\documentclass[]{aa}
\usepackage{graphicx}
\usepackage{txfonts}
\def\etal{et  al.\ }
\def\araa{{Ann.\ Rev.\ Astron.\ Ap.}}

\def\apj{ApJ}

\def\apjs{{ApJ\ Suppl.}}

\def\aa{{A\&A}}
\def\aap{{A\&A}}
\def\mnras{{MNRAS}}

\def\astroph{{Astro-ph}}

\def\keV {\rm keV}

\def\rc{{r_{\rm c}}}
\def \rci {r_{\rm ci}}
\def \rcut {R_{\rm cut}}

\def\ne{{n_{\rm e}}}

\def\nh{{N_{\rm H}}}

\def\kT{{\rm k}T}
\def\fgas{f_{\rm gas}}
\def\rs {r_{\rm s}}

\def\betamod{$\beta$-model}
\def \xmm {\hbox{\it XMM-Newton }}

\def \etal {et al.\ }

\def \xspec {\hbox{\sc xspec}}
\def \evigweight {\hbox{\sc evigweight}}
\def \asmooth{\hbox{\sc asmooth}}

\def \wabs {\hbox{\sc wabs}} \def \mekal {\hbox{\sc mekal}}

\newcommand{\hms}[3]{#1$^h$#2$^m$#3$^s$}
\newcommand{\dms}[3]{#1$^\circ$#2$'$#3$''$}

\begin{document}
\title{\xmm observation of the relaxed cluster A478: gas and dark
matter distribution from $0.01 R_{200}$ to $0.5 R_{200}$}

            \author{E. Pointecouteau \inst{1}, M. Arnaud \inst{1}, J.
            Kaastra \inst{2}, J. de Plaa \inst{2}}
\offprints{E. Pointecouteau, \email{pointeco@discovery.saclay.cea.fr}}

         \institute{Service d'Astrophysique du CEA , L'Orme des Merisiers,
Bat. 709, F-91191 Gif sur Yvette, France \\
         SRON, Sorbonnelaan 2, 3584 CA Utrecht,  The Netherlands \\
} \date{Received ; accepted } \abstract{ We present an \xmm
mosaic observation of the hot ($kT\sim6.5$~keV) and nearby ($z=0.0881$)
relaxed cluster of galaxies A478.  We derive precise gas density, gas
temperature, gas mass
and total mass profiles up to $12\arcmin$ (about half of the virial
radius  $R_{200}$).  The gas density profile is highly peaked towards
the center and the surface brightness profile is well fitted by a sum
of three $\beta$--models.
The derived gas density profile is  in excellent agreement, both in
shape and in normalization,  with the published Chandra density 
profile (measured
within $5\arcmin$ of the center).  Projection
and PSF effects on the temperature profile determination are thoroughly
investigated.
The derived radial temperature  structure is as expected for a cluster
hosting a cooling  core, with a
strong negative gradient at the cluster center.  The temperature rises
from $\sim2$~keV up to a plateau of $\sim6.5$~keV beyond $2\arcmin$ (i.e.
$r>208\rm{kpc}=0.1\, R_{200}$, $R_{200}=2.08$~Mpc being the virial radius).
 From the temperature profile
and the density profile and on the hypothesis of hydrostatic
equilibrium, we derived the total mass profile of A478 down to 0.01 and
up to 0.5 times the virial radius. We tested different dark matter
models against the observed  mass profile.  The Navarro, Frenk \& White
(\cite{navarro97}) model is significantly preferred to other models.
It leads to a total mass of $M_{200}=1.1\times 10^{15}$~M$_\odot$ for
a concentration parameter of $c=4.2\pm0.4$.  The gas mass
fraction  increases slightly with radius. The gas mass fraction  at a
density contrast of $\delta=2500$ is $\fgas=0.13\pm0.02$, consistent
with
previous results on similar hot and massive clusters.  We confirm the
excess of absorption in the direction of A478. The derived absorbing
column density   exceeds the 21~cm measurement by a factor of $\sim2$,
this excess extending well beyond the cool core region.  Through the
study of this absorbing component and a cross
correlation with infrared data, we argue that the absorption excess
is of Galactic origin, rather  than intrinsic to the cluster.
}

\authorrunning{Pointecouteau \etal }
\titlerunning{\xmm observation of the relaxed cluster A478}
\maketitle

%
%________________________________________________________________

\section{Introduction}

As nodes of large scale structure and thus places of dark matter
concentration, galaxy clusters can be used as powerful tools to test
theories
of structure formation.  The basic hierarchical scenarios based on
gravitation make the population of galaxy clusters a homologous
population of sources.  Their physical properties follow scaling laws
depending only on their redshift and mass, and their internal structures
are similar.

The exceptional capabilities of \xmm in terms of sensitivity and of
Chandra in term of spatial resolution allow us to characterize the gas
density and temperature profiles with  unprecedented accuracy.  For
a relaxed cluster, the hydrostatic equations can be used to derive the
underlying dark matter distribution, from the very central part of
clusters up to nearly the virial radius (David \etal~\cite{david01};
Allen \etal~\cite{allen01b}; Arabadjis, Bautz \& Garmire~\cite{abg02};
Allen, Schmidt \& Fabian~\cite{allen02a}; Pratt \&
Arnaud~\cite{pratt02},\cite{pratt03}; Lewis \etal~\cite{lewis03}; 
Buote \& Lewis~\cite{buote03}).
The observed clusters seem to have a cusped dark matter profile as
predicted by numerical simulations (Navarro, Frenk \&
White~\cite{navarro97}; hereafter NFW; Moore \etal~\cite{moore99};
hereafter MQGSL).  However, the central slope of the dark matter
profile and the
possible dispersion of the concentration parameter remain open issues.
Larger samples of high quality mass profiles are needed to further
assess these points.

In this paper we present the \xmm spectro-imaging observation of A478,
a massive, relaxed nearby cluster ($z=0.0881$ -- \cite{struble99}).
Detected in surveys (UHURU, HEAO-1, Ariel-V), this cluster is well
known in X-rays and its physical properties have been carefully
studied with previous X-ray observatories: EXOSAT (Edge \&
Stewart~\cite{edge91}), Einstein and Ginga (Johnstone
\etal~\cite{johnstone92}), ROSAT (Allen \etal~\cite{allen93}; White
\etal~\cite{white94}) and ASCA (\cite{markevitch98}, White \etal
\cite{white00}).  All those previous studies converge for what
concerns the overall temperature of the cluster, $kT \sim 6.8$~keV.
Recently, Sun \etal (\cite{sun03}) performed a high angular resolution
study of the central part of the cluster with Chandra.  They pointed
out the presence of an X-ray cavity in the very central part of the
cluster which is anti-correlated with the radio lobes.

Here we focus on the characterization of the gas and dark
matter distribution of A478.  In a companion paper de Plaa \etal (in
prep.) present a detailed spectroscopic study of the metal
abundances and their distribution within A478's core based on EPIC and
RGS
data.  We present the observation and the different data processing
steps in Sect.~\ref{sec:data}.  In Sect.~\ref{sec:morpho} we briefly
discuss the cluster morphology.  In Sect.~\ref{sec:ne} we analyze the
surface brightness profile and derive the gas density profile.
Spatially resolved spectroscopic analysis is presented in
Sect.~\ref{sec:spec}, where we also discuss the temperature and
absorption
profiles.  In Sect.~\ref{sec:mass}, we present the resulting total
mass and gas mass fraction profiles of A478 and we discuss the shape of
the dark
matter profile according to our observational results.

Throughout this paper,  we use $H_0=70$~km Mpc$^{-1}$ s$^{-1}$,
$\Omega_m=0.3$ and
$\Omega_{\Lambda}=0.7$. In such a cosmological framework, at the cluster
redshift ($z=0.0881$) $1$\arcmin~$=99$~kpc.

%__________________________________________________________________

\section{\xmm observations and data processing \label{sec:data}}

\subsection{Observations}
A478 was observed with \xmm  during revolution 401 for a total exposure
time of 126~ks with the EMOS camera and 122~ks with the EPN camera.
For the first 70~ks two CCDs ($CCD \#4$ and $\#7$) of the EMOS1 camera
were not operating.  An offset observation, centered at
$\alpha=$\hms{04}{12}{35},
$\delta=$\dms{10}{15}{45}, $17\arcmin$ South West of the cluster center,
was performed in revolution 401 for 38 ks (EMOS camera) and 34 ks
(EPN camera).  All observations were performed in EXTENDED FULL FRAME
mode and using the THIN1 filter.  We used the calibrated events list
produced by the \xmm SOC pipeline and processed them with the SAS
(v5.4.1).  We also used the blank-sky event files accumulated by Lumb
\etal~(\cite{lumb02}).

\subsection{Event list screening}

In this work, we only kept events with pattern 0 to 12 from EMOS data,
pattern 0 from EPN data and flag equal 0 for both detectors.

First the event list for each camera and each observation was
filtered for periods of high background due to soft proton flares.
Visual inspection of the light curve in a source-free energy band
($[10-12]~\keV$ for EMOS and $[12-14]~\keV$ for EPN) revealed long
periods of high background.  These were excluded.  We then
fine-tuned the flare cleaning by using a ``3$\sigma$ clipping''
selection of good time intervals as described in Pratt \& Arnaud
(\cite{pratt03}).  The Poisson fit of the light curve histogram
provides a 3$\sigma$ threshold above which corresponding frames are
discarded.  To excise all flaring periods, this clipping method was
applied in different energy bands: $[2-5]~\keV$, $[5-8]~\keV$
$[8-10]~\keV$ and $[10-12]~\keV$ for EMOS; $[5-8]~\keV$,
$[8-10]~\keV$, $[10-12]~\keV$ and $[12-14]~\keV$ for EPN. This quite
conservative choice was made to avoid any low energy flares that are
present in some observations (Pratt \& Arnaud \cite{pratt02}).

The remaining exposure times after cleaning are $48.0$~ks,
$40.9$~ks and $37.3$~ks for the central pointing and the EMOS1, EMOS2
and EPN camera respectively.  The corresponding times for the offset
pointing are $13.1$~ks for the EMOS camera and $11.0$~ks for the
EPN camera.

        \begin{figure*}[!thh]
\centering
\includegraphics[width=18cm]{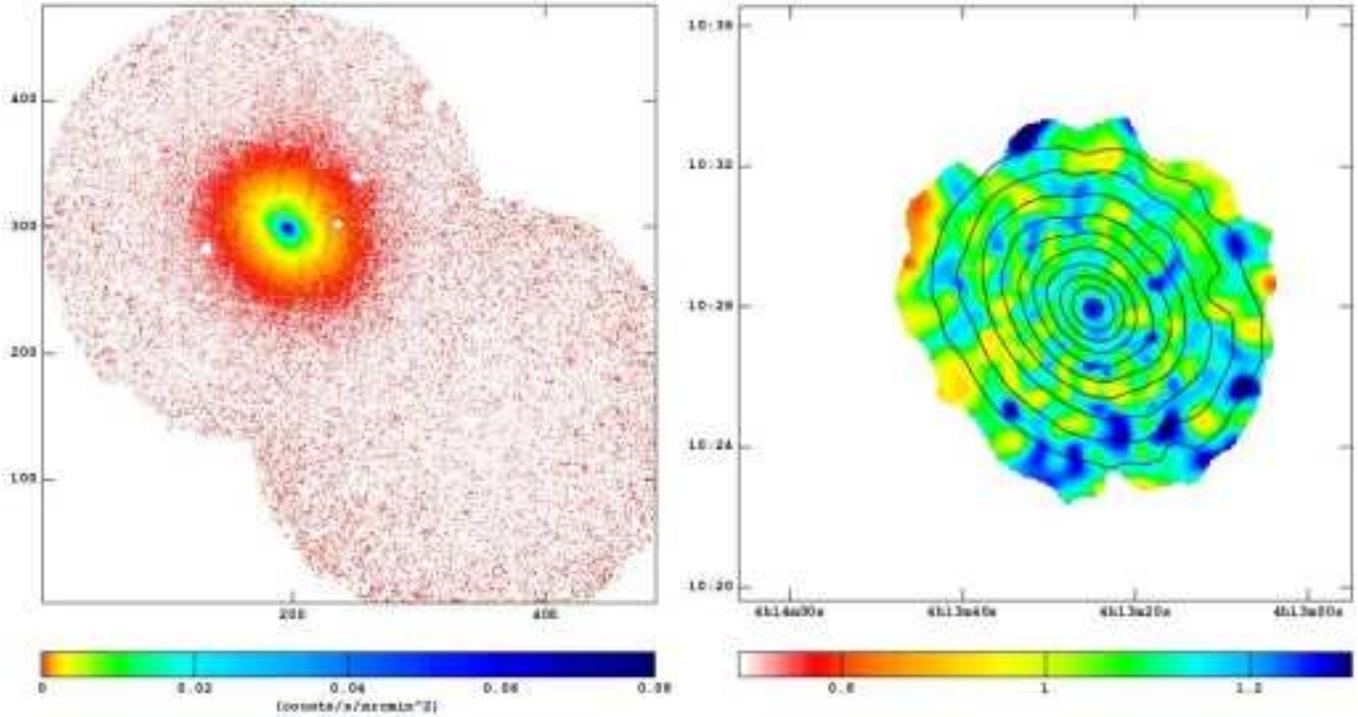}%
\vspace*{-0.5em} \caption{(left panel) Total (EMOS+EPN) \xmm mosaic
image of
A478 in the $[0.3-7]~\rm{keV}$ energy band (left panel).  The image is
corrected for vignetting and exposure, and background subtracted using
blank field data.  Right panel: hardness ratio map (ratio of the count
rates in the $[1.7-7]~\keV$ and in the $[0.3-1.7]~\keV$ energy bands).
The smoothed contours of the mosaic image have been overlaid.
\label{fig:f1}}
         \end{figure*}

\subsection{Extraction of imaging data and spectra}

For each camera, we merged the cleaned event lists of the two
observations into a single event list.  We only selected events
falling within a $15\arcmin$ circular region centered on the detector
optical axis.
The sky pixel coordinates of each event of the offset pointing were
re-projected onto the sky pixel grid of the central pointing.

Scientific products (spectra, images, surface brightness profiles) can
thus be extracted in a single step from the merged events list,
simplifying the analysis.  However, the exposure time can vary
strongly in a given extraction region and this has to be taken into
account in estimating count rates.  The extraction method we used is
fully described in Appendix~\ref{sec:app}.  All the products are
corrected for vignetting in this extraction process.  The
vignetting correction is based on the photon weighting method
described in Arnaud \etal~(\cite{arnaud01}), the weight coefficients
being tabulated in the event list with the SAS task \evigweight.

\subsection{Background subtraction \label{sec:bkg}}

For each camera, a clean background event list was extracted from the
corresponding blank-field data using the same filtering criteria
(pattern and flag selection, 3$\sigma$ flare clipping) as for the
observations.  Cast-background files were then generated for each
pointing by applying the aspect solution of the observation to the
background dataset.  The correct exposure time was computed for each
blank field event list.

To estimate the difference of particle background level
between each observation and the corresponding blank field data, we
computed the ratio of the total count rates in the high energy band
($[10-12]~\keV$ and $[12-14]~\keV$ respectively for EMOS and EPN data).
    As
the expected average temperature of this bright cluster is quite high
($\sim 6.8~\keV$) we excluded a circular central region
$(\theta<6\arcmin)$ to avoid any contamination due to the cluster.

Blank field products have to be normalized by this ratio, when used
as a background for the corresponding observation products.  The
normalizations are slightly different from one pointing to the other.
To take that normalization into account, we multiplied the weight
coefficients in each cast background file by the corresponding
normalization factor.  Vignetting-corrected blank field counts are
then automatically `normalized'.  These cast background files of the two
pointings are then  merged as for the observations.  Product
extraction is then the same (see Appendix~\ref{sec:app}).

For the EPN data, we generated a list of out-of-time events (OoT
hereafter)
to be treated as an additional background component.  An OoT event
occurs
when a photon is detected during the read-out process.  The current
observing mode (Extended Full Frame) minimizes the effect of OoT 
to 2.3\%.  The OoT event list was processed similarly to the observation
EPN event files.

The background subtraction (for spectra and surface brightness
profiles) is performed as described in full detail in Arnaud
\etal~(\cite{arnaud02b}).  It is a two-step procedure, which insures a
correct cosmic X-ray background (CXRB) subtraction, even when the
local CRXB is different from the average CRXB in blank field data.  In
a first step, for each product extracted from the merged observation
event list, an equivalent product is extracted from the corresponding
merged blank-field file and subtracted from it.  For EPN the
OoT data are also subtracted.  This first step allows us to remove the
particle background.  However, it may over(under) subtract the CRXB if
the CRXB in the observation region is smaller(larger) than the average
value in the blank field observation.  The residual CXRB (i.e. the
difference between the CXRB in the A478 field and in the blank field)
is then estimated by using blank field subtracted data in the region
free of cluster emission ($\theta>16\arcmin$ from the cluster center
-- The cluster is significantly detected in the background subtracted
surface brightness profile up to 12-13\arcmin).  This residual CRXB is
subtracted in a second step from each EMOS and EPN product.  In our
case the residual is negative in the $[0.3-3]~\keV$ energy band.  The
residual count rate summed over the three detectors is
-0.68~counts/s/arcmin$^2$, which represents 23\% of the total
background count rate in this energy band.  Beyond 3~keV the residual
background ($r>16\arcmin$) spectrum is consistent with zero. Therefore
the double subtraction beyond 3~keV will only contribute increasing
the noise level in each channel. To minimize errors, the double
subtraction is thus applied only to data in the energy band
$[0.3-3]~\keV$. We nevertheless check (on the global spectrum) that
the best fit values remain the same if a full double subtraction is
applied.

\subsection{Point source exclusion}

Starting from the output of the SAS detection source task, we made a
visual selection on a wide energy band EMOS \& EPN image (extracted from
the merged event lists) of point sources in the FOV. Events from
these regions were excluded directly from each merged event
list.  We generated corresponding mask mosaic images, which were then
used
to compute the surface of each extraction region.

%__________________________________________________________________

\section{Cluster morphology \label{sec:morpho}}

The mosaic count rate image (EMOS+EPN) in the $[0.3-7]~\keV$ energy
band is
presented in Fig.~\ref{fig:f1}, together with the hardness ratio map
computed from the ratio of the images in the $[1.7-7]~\keV$ and
$[0.3-1.7]~\keV$ energy bands.  The images in various energy bands are
vignetting corrected and background subtracted using the corresponding
blank field image and the OoT image in case of EPN.  To generate the
hardness ratio map, the hard band image was first adaptively smoothed
(with the task \asmooth).  The soft band image was then smoothed using
the same smoothing template as was created for the hard band image.  The
cluster morphology is regular and the hardness ratio map does not
exhibit any peculiar feature, reinforcing the assessment that A478 is
a very relaxed cluster.

A478 has an elliptical shape.  From optical and ROSAT/PSPC and HRI
data White \etal~(\cite{white94}), derived an ellipticity $\epsilon$
varying
between $1.2$ and $1.4$ within the central $(\theta < 100\arcsec)$
region.
We fitted a 2D $\beta$-model to the EMOS1+EMOS2 image in the
$[0.3-2]~\keV$
energy band, within the $\theta < 10\arcmin$ region.  We derived a
consistent value of the ellipticity, $\epsilon=1.22$.  The quality of
the fit is poor, however: the residual image shows a strong excess at
the cluster center position, as expected for a strong cooling flow
cluster.

Despite its slightly elliptical shape, in the remainder of this work we
assume spherical symmetry and use circular annuli to extract the surface
brightness profile and spectra.  Pratt \& Arnaud~(\cite{pratt02})
showed in
the case of A1413, a cluster of a higher ellipticity of
$\epsilon=1.4$, that this has negligible impact on the derived
temperature and mass profiles.

\section{Gas density radial profile}
\label{sec:ne}

\subsection{Surface brightness profile \label{spapr}}
\label{sec:sx}

\begin{figure}[ttt]
\centering
\hspace{-3em}%
\includegraphics[width=9cm]{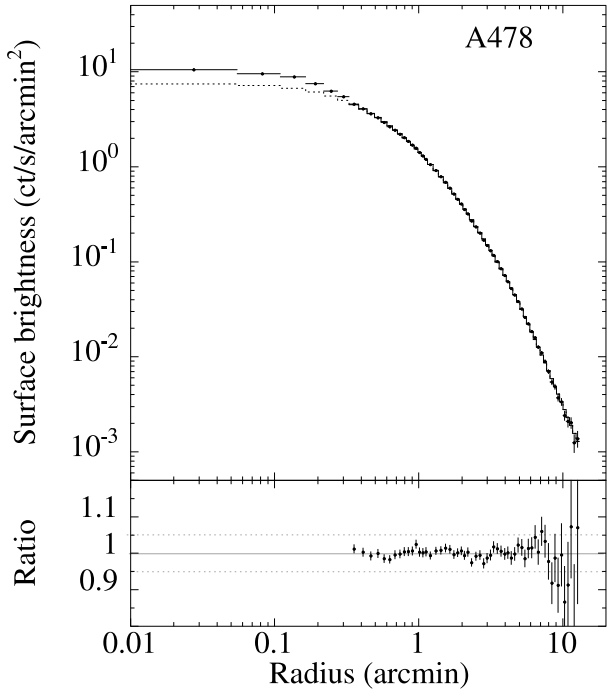}
\vspace{-5em}
\caption{A478
  \xmm (EMOS+EPN) surface brightness profile in the $[0.3-2.0]~\keV$
  band.  The profile is background subtracted, corrected for
  vignetting and for the radial variations of the emissivity factor
  (see text).  The best fit KBB gas density model (Eq.~\ref{eq:KBB}),
  fitted over the $\theta > 0.33\arcmin$ region is over-plotted as a
  solid line.  The dotted line is the extrapolation of this model in
  the central region.  Bottom panel: ratio between the data and the
  model. The dotted lines indicate the $\pm$5\% level.
         \label{fig:f2}}
          \end{figure}

          \begin{figure}[ttt]

\centering
\hspace{-2em}%
\includegraphics[width=9cm]{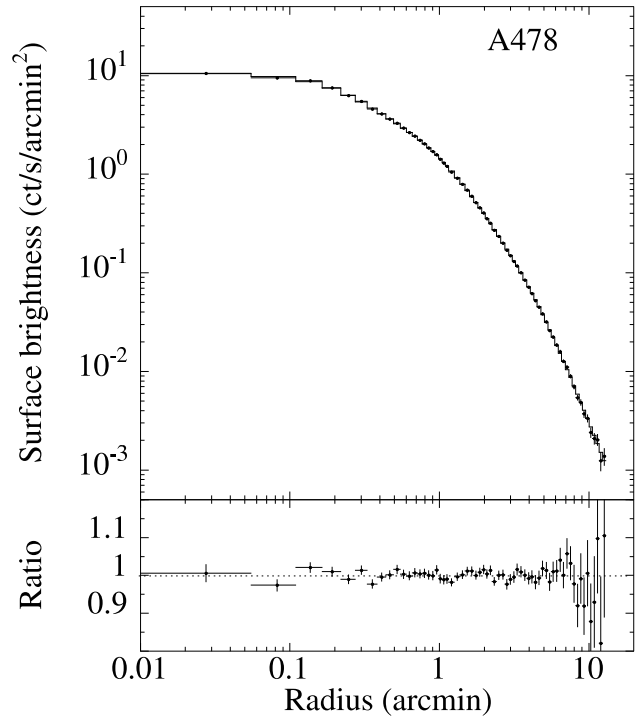}
\vspace*{-11em}
\caption{Same as
  Fig.~\ref{fig:f2} with the best fit BBB model (Eq.~\ref{eq:BBB})
  fitted over the full radial range.
         \label{fig:f3}}
          \end{figure}

We computed a background-subtracted vignetting-corrected radial surface
brightness profile for each detector in the $[0.3-2]~\keV$
energy band.  The width of the radial bins was $3.3\arcsec$.  The
profiles for the three detectors were added into a single profile,
which was rebinned to reach a significance level of at least 3$\sigma$
in each radial bin.  The cluster emission is detected up to $13\arcmin$.

The emissivity in the considered energy band varies slightly with
radius, due to the radial gradients in the hydrogen column density
along the
line of sight ($\nh$), temperature ($\kT$) and metallicity ($Z$)
(See Sect.~\ref{sec:speraw}).  The $\nh$, $\kT$ and
$Z$ values were extrapolated at each radius of the surface brightness
profile by fitting the observed $\nh$, $\kT$ and $Z$ profiles (See
Sect.~\ref{sec:speraw}) with a 3 degree polynomial, an empirical
temperature profile as described in Allen \etal~(\cite{allen01b}) and
a lognorm law respectively.  The corresponding emissivity profile
(with errors) was estimated using an absorbed redshifted \mekal\ model,
convolved with the instrument response.  Its radial variation is
mainly dominated by the variations of $\nh$.  The surface brightness
profile was then divided by the emissivity profile normalized to its
value at large radii.  The errors on the emissivity were propagated to
the corrected surface brightness profile.

\subsection{Gas density profile modeling \label{sec:nefit}}

The corrected surface brightness profile (presented in
Fig.~\ref{fig:f2}) is proportional to the emission measure profile,
$EM(r)$, and
can be fitted directly using various parametric models of the gas
density profile, $\ne(r)$.  The corresponding emission measure models
were convoluted with the \xmm\ PSF (Ghizzardi
\etal~\cite{ghizzardi01},~\cite{ghizzardi02}) and binned in the same
way as the observed profile.

As expected, a standard \betamod\ provides an unacceptable fit, the
data showing a strong excess in the center compared to the model.
Progressively cutting the central region decreases the reduced
$\chi^2$.  The fit becomes acceptable ($\chi_{\rm red}^2\sim 1$) for a
cut-out radius of $\theta_{\rm cut} \sim 2.4\arcmin$ with a best fit
$\beta$ value of $0.79\pm0.01$ ($1\sigma$ errors).  The reduced
$\chi^{2}$ keeps decreasing with increasing $\theta_{\rm cut}$ until
it stabilizes to a value of $\chi_{\rm red}^2\sim 0.7$ for
$\theta_{\rm cut} > 3\arcmin$.  There is an indication that $\beta$
increases with increasing $\theta_{\rm cut}$, but the effect is
marginally significant: for instance we obtained $\beta = 0.85\pm0.05$
for $\theta_{\rm cut}= 3.9 \arcmin$.

We then considered the alternative parametrisations of the gas
density profile
proposed by Pratt \& Arnaud~(\cite{pratt02}) for cases where a central
excess is seen.  We fitted the entire profile with a cusped model,
similar to
the NFW profile (their AB model) and a double isothermal \betamod\
(their BB model).  Both models fail to account for the data, although
formally the latter provides  a better fit than the former: the reduced
$\chi^2$ are respectively $5.5$ and $4.2$.  We then tried a
generalized double \betamod\ (their KBB model).  This model allows a
more centrally peaked gas density profile in the core than the BB
model and is defined by:
\begin{equation}
\begin{array}{lllll}
r < \rcut &\ne(r)& = &\ne(0) &\left[ 1 +
\left(\frac{r}{\rci}\right)^{2\xi}\right]^{-\frac{3 \beta_{\rm
i}}{2\xi}} \\

r > \rcut &\ne(r)& =& N &\left[ 1 + \left(\frac{r}{\rc}\right)^2
\right]^{-\frac{3 \beta}{2}}\\
        \end{array}
\label{eq:KBB}
\end{equation}
\noindent where $\xi$, $\rci$, $\rcut$, $\rc$ and $\beta$ are free
parameters, the parameters $N$ and $\beta_{\rm i}$ being related to
them so that both the density distribution and its gradient are
      continuous across $\rcut$.

This model provides an excellent fit to the data, but only if the very
central part is excluded from the fit (see Fig.~\ref{fig:f2}).
Fitting the $(\theta > 0.33\arcmin \sim 32~{\rm kpc})$ region gives a
reduced $\chi^{2}$ of $\chi_{\rm red}^2\sim 0.92$ for $\xi = 0.42$,
$\rci=0.99\arcmin$, $\rcut=4.5\arcmin$, $\rc=2.5\arcmin$ and
$\beta=0.81$.  A clear excess is observed when extrapolating this
model in the central part (see Fig.~\ref{fig:f2}).

         \begin{figure}[!ttt]
\centering
\hspace{-2em}%
\includegraphics[width=9cm]{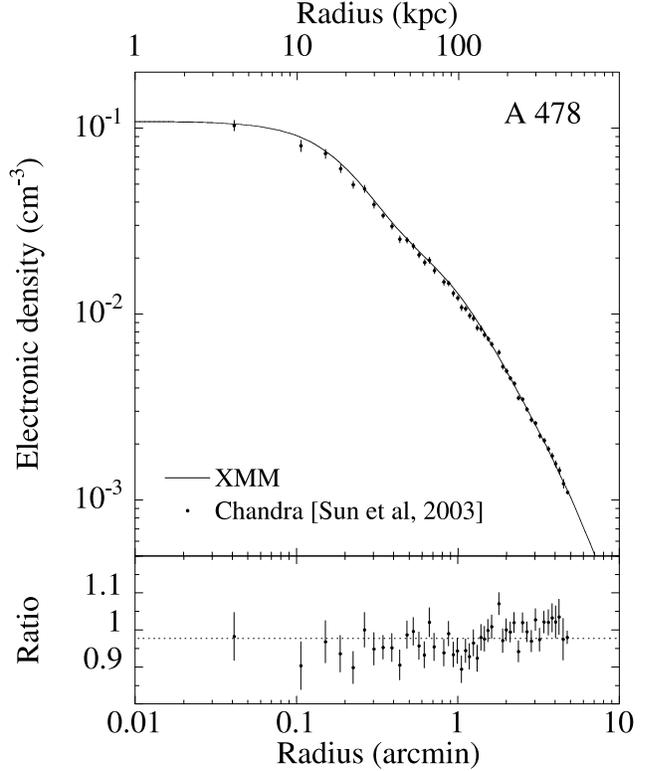}
\vspace*{-3em}
     \caption{Comparison
       between the Chandra gas density profile derived by Sun
       \etal~(\cite{sun03}) and the \xmm best fit BBB gas density
       model (as plotted in Fig.~\ref{fig:f3}).  Bottom panel: ratio
       between the Chandra data and this model.
         \label{fig:f4}}
          \end{figure}

We then tried a 3 component model.  We went back to direct
parametrisation of the emission measure profile (rather than the
density profile) for simplicity.  We considered the sum of three
$\beta$-models (hereafter BBB model).
\begin{equation}
\begin{array}{lll}
        EM(r) &=& EM_{1}(r) + EM_{2}(r) + EM_{3}(r)\\
        EM_{i}(r)& = & EM_{i}(0)\left[ 1 +
\left(\frac{r}{\rc_{i}}\right)^2 \right]^{-3 \beta+1/2}
\end{array}
\label{eq:BBB}
\end{equation}
A common $\beta$ value is assumed to insure a smooth (single power law)
behavior at high radii.  The model has thus $7$ free parameters. The
corresponding density profile is:
\begin{equation}
\begin{array}{lll}
        \ne(r)& =& \sqrt{\ne_{1}^{2}(r) + \ne_{2}^{2}(r) + \ne_{2}^{2}(r)}
\end{array}
\end{equation}
where each $\ne_{i}(r)$ profile is the density \betamod\ profile
corresponding
to the emission measure profile $EM_{i}(r)$.

\begin{table*}[!t]
           \caption[]{A478 global properties from previous studies}
              \label{tab1}
\begin{flushleft}
$$
         \begin{tabular}{lcccccc}
         \hline
         \hline
         & Exosat~$^{\mathrm{a}}$ & Ginga/Einstein~$^{\mathrm{b}}$ &
ROSAT/Ginga~ $^{\mathrm{c}}$ & ASCA~ $^{\mathrm{d}}$ &
Chandra~$^{\mathrm{e}}$ & XMM~ $^{\mathrm{f}}$ \\
         \hline
         $kT$ (keV) & 6.8$^{+1.1}_{-1.0}$ & 6.84$^{+0.2}_{-0.24}$ &
6.56$^{+0.14}_{-0.14}$ & 6.58$^{+0.26}_{-0.25}$ &
7.18$^{+0.11}_{-0.11}$ & 6.17$^{+0.12}_{-0.06}$\\
         $\nh$ ($10^{21}$cm$^{-2}$) & 1.1$^{+1.9}_{-0.7}$ &
3.6$^{+0.2}_{-0.2}$ & 2.49$^{+0.12}_{-0.09}$ & 3.0~$^{\mathrm{g}}$ &
2.59$^{+0.03}_{-0.03}$ & 2.66$^{+0.02}_{-0.03}$ \\
         $Z$ ($Z_\odot$) & 0.27$^{+0.14}_{-0.16}$ & 0.37$^{+0.04}_{-0.04}$ &
0.39$^{+0.04}_{-0.04}$ & 0.31$^{+0.03}_{-0.03}$ &
0.37$^{+0.02}_{-0.02}$ & 0.32$^{+0.01}_{-0.01}$ \\
         \hline
         \hline
              \end{tabular}
$$
\end{flushleft}
$^{\mathrm{a}}$Edge \& Stewart~\cite{edge91},
$^{\mathrm{b}}$Johnstone \etal~\cite{johnstone92},
$^{\mathrm{c}}$Allen \etal ~\cite{allen93} and White
\etal~\cite{white94},
$^{\mathrm{d}}$White~\cite{white00}, $^{\mathrm{e}}$Sun
\etal~\cite{sun03}, $^{\mathrm{f}}$ This work.  \\
$^{\mathrm{g}}$~fixed parameter.\\
Note: all studies include the cooling core regions.
        \end{table*}

This BBB model provides an excellent fit to the data over the whole
radial
range: the reduced $\chi^{2}$ is $\chi_{\rm red}^2\sim 0.90$ for $56$
d.o.f. This best fit model is plotted in Fig.~\ref{fig:f3}, together
with the ratio between data and model.  The best fit parameters are
$\beta=0.84^{+0.04}_{-0.03}$, $\rc_{1} =
0.25\arcmin\pm0.02\arcmin$, $\rc_{2} =
1.12\arcmin\pm0.07\arcmin$, $\rc_{3} =
3.3\arcmin\pm0.3\arcmin$ ($90\%$ errors).  The best fit central
density is $\ne(0) =0.109 {\rm cm^{-3}}$, with a relative normalization
of the second and
third components of $(\ne_{2}(0)/\ne(0))^{2} = 5.40~10^{-2}$ and
$(\ne_{2}(0)/\ne(0))^{2}=1.32~10^{-3}$, respectively.

We further compared this best fit density profile with the profile
obtained by Sun \etal~(\cite{sun03}) from a deprojection of the
Chandra ACIS-S3 surface brightness (see Fig.~\ref{fig:f4}).  The
Chandra profile is determined out to $\sim 5\arcmin$.  There is a good
agreement between the two profiles, both in shape and in
normalization.  Simply adjusting the overall normalization of the \xmm
best fit model
to the Chandra data gives already a reduced $\chi^{2}$
of  $1.3$ with residuals between model and data less than $5\%$. The
normalization is $2\%$ lower than the \xmm value,  corresponding to
$\sim 4\%$ discrepancy in X--ray
flux, well within the discrepancy of $\sim \pm 5\%$ between the two
instruments found by Snowden~(\cite{snowden02}).  Although the
best fit XMM model does not fit perfectly the Chandra profile shape,
slightly adjusting the parameters provides an acceptable fit.  We kept
the $\beta$ and outer core radius values to their best fit XMM values,
as the external shape of the density profile is not well constrained
by Chandra data.  We obtained a reduced $\chi^{2}$ of $1.0$ for
$\rc_{1} = 0.26\arcmin$ and $\rc_{2} = 1.18\arcmin$,
consistent with the XMM $90\%$ confidence range.  The relative
normalizations of the three components are marginally inconsistent.
Nevertheless, taking into account that the effect of the Chandra PSF
is negligible, the good agreement between XMM and Chandra central core
radius indicates that the PSF modeling we have used to fit the \xmm
data is basically correct.

The XMM best fit BBB model is thus used in the following to
estimate the cluster gas and total mass profiles and to correct the
temperature profile for PSF and projection effects.  The total mass
profile depends on the logarithmic slope of the density profile.  To
estimate the systematic uncertainties in the mass estimates we will
also consider the density logarithmic slopes derived from the BBB
model best fitting the Chandra profile.  The differences are small
however, in the range $0.2 - 4.5\%$.  Finally we would like to
emphasize that the BBB model functional form must only be viewed as a
convenient parametric representation of the gas density profile.  It
has no particular physical ground and must not be over-interpreted
(e.g in terms of three distinct gravitational systems in the cluster).
%__________________________________________________________________

\section{Temperature and absorption profiles \label{sec:spec}}

Throughout the analysis, the spectra are binned to reach a
significance level of at least 3$\sigma$ in each bin.  We used \xspec~
to fit the data with an absorbed redshifted thermal model
({\wabs(\mekal)}).  Due to larger calibration uncertainties in the
instrument response below the O edge we only fitted the spectra above
$E =0.6~\keV$.  We used the following response matrices:
m1\_thin1v9q20t5r6\_all\_15.rsp (EMOS1),
m2\_thin1v9q20t5r6\_all\_15.rsp (EMOS2) and epn\_ef20\_sY9\_thin.rsp
(EPN, created in December 2002).  Unless otherwise stated, the
relative normalizations of the EPN and EMOS spectra were left free when
fitted simultaneously.

\subsection{Global spectrum analysis \label{speall}}

We first extracted the cluster EMOS1, EMOS2 and EPN spectra within a
circular region of $10\arcmin$.  Fitting simultaneously the EMOS and
EPN spectra, we obtained a redshift of $z=0.0868\pm 0.0004$.
This value is significantly smaller than the optical value
($z=0.0881\pm0.0009$).  An investigation of the variation of
$\chi^{2}$ with $z$ revealed two local minima, one at the optical
redshift location and one at $z= 0.079$.  Independent spectral fits
of EMOS and EPN spectra clarify the issue.  The EMOS best fit value is
$z=0.0889\pm 0.0010$, perfectly consistent with the optical
value.  On the other hand, the EPN best fit redshift is
$z=0.0793\pm 0.0007$ and corresponds to the second minimum.
The redshift difference from the optical value corresponds to an
energy shift of $\Delta E = +50\pm4~{\rm eV}$ with respect to
the expected iron line position.
         \begin{figure}[!t]
         \hspace*{-1em}%
         \includegraphics[width=9cm]{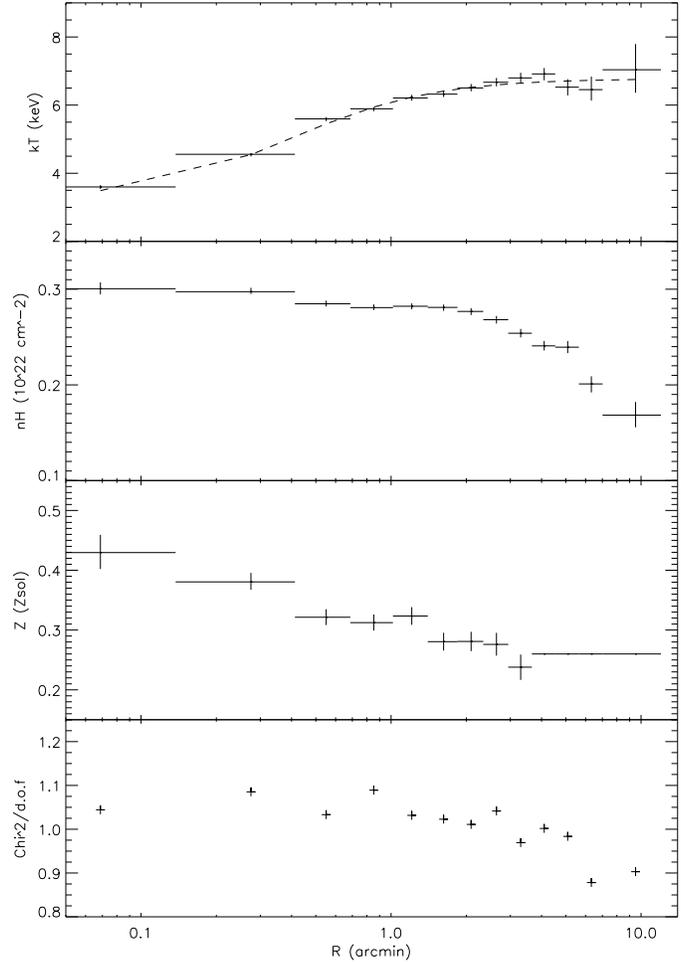}
         \caption{Result of the radial spectral fitting. From top to
           bottom the temperature ($kT$), Galactic absorption ($\nh$),
           metallicity ($Z$) and reduced $\chi^{2}$ for the best fit
           model.  Errors are $1\sigma$.  For the temperature profile
           the best fit model using Eq.~(\ref{eq:fitkt}) is over-plotted
           as a dashed line.
         \label{fig:f5}}
          \end{figure}

We thus checked for a possible gain problem in the EPN data.  The
non-background subtracted spectrum shows strong Al-K, Ni-K and Cu-K
fluorescence lines, which can be used for this purpose.  We fitted the
spectrum extracted in the $6\arcmin < \theta < 12\arcmin$ region in a
restricted energy range around these lines with a power law plus
Gaussian line(s).  The fitted Ni-K and Cu-K line energies are
significantly higher than expected: $E=7517\pm6~{\rm eV}$ and
$E=8081\pm1.5~{\rm eV}$ respectively, to be compared to the expected
values of $7477~{\rm eV}$ and $8047~{\rm eV}$.  However, the
discrepancies, $\Delta E = 40\pm6~{\rm eV}$ and $\Delta E=34\pm2~{\rm
eV}$ respectively, are smaller than observed for the cluster iron line.
Furthermore we cannot simply add a constant offset to the energy
scale: the centroid energy of Al-K, $E=1489^{+4}_{-3}~{\rm eV}$,
is consistent with the expected value of $1487~{\rm eV}$.  Any linear
gain correction based on the position of these fluorescence lines
would thus be insufficient to bring the EPN redshift determination into
agreement with the optical value.

We thus did not try to define and apply a gain correction
In principle, the EPN gain
uncertainty could affect our results.  To assess this point, we fitted
the EPN spectrum both fixing the redshift at the optical value and
letting this parameter free.  The derived $\nh$ values are the same,
the best fit temperatures differ by $0.10~\keV$, similar to the
statistical uncertainty.  As expected the main impact is on the
derived abundance: for a free redshift $Z=0.34\pm0.02~{\rm
Z_{\odot}}$, only marginally consistent with the value,
$Z=0.30\pm0.02~{\rm Z_{\odot}}$, obtained fixing the redshift at the
optical value.  When the EPN spectrum is fitted simultaneously with
the EMOS spectrum, these discrepancies are even smaller.  The
abundance difference is two times less and the temperature is the same.

As we are mostly interested in the temperature information, we can
neglect the gain uncertainty and choose to fix the redshift to the
optical value, in all the following
analysis.  The overall cluster  parameters are:
$\nh=2.66_{-0.03}^{+0.02}\times 10^{21}$~cm$^{-2}$,
$\kT=6.17_{-0.06}^{+0.12}$~keV and $Z=0.32\pm 0.013$~Z$_{\odot}$
($\chi_{\rm red}^2 = 1.02$ -- 90\% confidence level).

\subsection{Annular spectra analysis
\label{sec:speraw}}

We extracted background-subtracted, vignetting-corrected spectra in 13
concentric annuli centered at the peak of the X-ray emission.  The
annuli were defined to have about the same number of counts per bins
(except the outermost annulus).

The EMOS1, EMOS2 and EPN spectra of each annulus were simultaneously
fitted with a \wabs(\mekal) model.  The resulting $\nh$, $\kT$, and
abundance profiles are plotted in Fig.~\ref{fig:f5}.  The definition
of the annuli and the best fit parameters are gathered in
Table~\ref{tab2}.  We also checked that the temperature profiles
obtained by fitting the EMOS and EPN spectra independently are
consistent within the error bars.

The temperature profile shows a clear drop towards the center.  It is
well fitted by the analytical formula proposed by Allen \etal
(\cite{allen01b}):
\begin{equation}
T(r)=T_0+T_1\left[\frac{(r/\rc)^\eta}{1+(r/\rc)^\eta}\right]
\label{eq:fitkt}
\end{equation}
with $T_{\rm 0}= 3.26~\keV$ and $T_{\rm 1}=3.52~\keV$, $\rc =
0.396\arcmin $ and $\eta=1.52$ ($\chi^{2}=12.7$ for 13 degrees of
freedom).  The best fit model is plotted in Fig.~\ref{fig:f5}. To fit
the
observed profile with such a formula, we had to assign a radius to
each annulus temperature.  Following the prescription of Lewis \etal
(\cite{lewis03}), we used the weighted effective radius of each annulus,
defined as:
\begin{equation}
r_{\rm i} = \left[(r_{\rm out_{i}}^{3/2} + r_{\rm
in_{i}}^{3/2})/2 \right]^{2/3}
\label{eq:rw}
\end{equation}
rather than the mean radius.  We checked that the best fit temperature
profile then becomes insensitive to the binning choice.
We regrouped the first two annuli and then the next two 
annuli and re-ran an isothermal fit to the spectra of  those new larger
annuli.
The resulting profiles were fitted again with Eq.~(\ref{eq:fitkt}), using
the weighted effective radii. We obtained the same best  fit profile
as with the original binning.
This is not the case if we use instead the mean radius of each annulus.

The temperature profile shows a strong gradient towards the center,
whereas we
recall that the surface brightness profile is very peaked.  This
temperature profile is thus likely to be affected by both PSF and
projection effects.  These effects will be analyzed in
Sect.~\ref{sec:specor}.  However, the overall $\nh$ value is
significantly
higher than the 21 cm value and its radial profile is not flat.  As
the $\nh$ and $\kT$ determination are not independent, we will first
discuss our absorption results.

        \begin{table}[!h]
           \caption[]{Radial spectral fitting -- best fit values and
associated error bars (90\% confidence level) }
              \label{tab2}
\begin{flushleft}
\vspace*{-1em}
$$
         \begin{tabular}{ccccl}
         \hline
         \hline
         $R_{out}$~$^{\mathrm{a}}$ & $\nh$ & $kT$ & $Z$ & $\chi_{\rm
         red}^{2}$ (d.o.f)  \\
         (arcmin) & ($10^{21}$cm$^{-2}$) & (keV) &  ($Z_\odot$)  \\
         \hline
0.14 & 3.00$\pm$0.10 & 3.60$\pm$0.10 & 0.43$\pm$0.05 & 1.04 (1140)\\
0.41 & 2.97$\pm$0.05 & 4.54$\pm$0.07 & 0.38$\pm$0.02 & 1.08 (1979) \\
0.69 & 2.85$\pm$0.05 & 5.60$\pm$0.10 & 0.32$\pm$0.02 & 1.03 (2111)\\
1.02 & 2.81$\pm$0.05 & 5.88$\pm$0.11 & 0.31$\pm$0.02 & 1.09 (2135) \\
1.40 & 2.82$\pm$0.05 & 6.21$\pm$0.14 & 0.32$\pm$0.02 & 1.03 (2052)\\
1.84 & 2.81$\pm$0.05 & 6.32$\pm$0.16 & 0.28$\pm$0.02 & 1.02 (1937)\\
2.34 & 2.77$\pm$0.06 & 6.50$\pm$0.17 & 0.28$\pm$0.03 & 1.01 (1926)\\
2.94 & 2.68$\pm$0.06 & 6.67$\pm$0.20 & 0.28$\pm$0.03 & 1.04 (1736)\\
3.66 & 2.54$\pm$0.07 & 6.80$\pm$0.25 & 0.24$\pm$0.04 & 0.97 (1559)\\
4.54 & 2.41$\pm$0.08 & 6.91$\pm$0.30 & 0.26$^{\mathrm{b}}$ & 1.00
(1405)\\
5.64 & 2.39$\pm$0.10 & 6.53$\pm$0.39 & 0.26$^{\mathrm{b}}$ & 0.98
(1153)\\
7.01 & 2.01$\pm$0.14 & 6.46$\pm$0.58 & 0.26$^{\mathrm{b}}$ & 0.88
(843)\\
12.0 & 1.68$\pm$0.22 & 7.04$\pm$1.29 & 0.26$^{\mathrm{b}}$ & 0.90
(443)\\
         \hline
         \hline
              \end{tabular}
          $$
\end{flushleft}
$^{\mathrm{a}}$ External radius of the annuli. $^{\mathrm{b}}$ fixed
parameter.

        \end{table}

\subsection{The absorption profile \label{sec:nh}}

Our best fit overall value for the Galactic absorption,
$\nh=2.66_{-0.03}^{+0.02}\times 10^{21}$~cm$^{-2}$, is nearly two
times the 21 cm value of $\nh=1.53\times 10^{21}$~cm$^{-2}$ (Dickey
\& Lockman~\cite{dickey90}).  Such an  excess absorption was found in
all previous X-rays studies of A478 and our derived value is only
marginally higher than the value derived from ROSAT/Ginga and Chandra
data (see Table~\ref{tab1}).

The radial $\nh$ profile that we obtained
exhibits a clear gradient ranging from $3\times 10^{21}$~cm$^{-2}$ in
the central regions to $1.7\times 10^{21}$~cm$^{-2}$ at $12\arcmin$.
This gradient is consistent with the
Chandra gradient measured by Sun \etal (\cite{sun03}): from a
central value of $2.9\pm 0.1 \times 10^{21}$~cm$^{-2}$ down to
$2.4\pm 0.1 \times 10^{21}$~cm$^{-2}$ for the last bin at $\sim
5\arcmin$ (perfectly consistent with the \xmm value, see
Table~\ref{tab2}).

This excess of absorption seen in A478 and other cooling flow clusters
was interpreted in previous studies (e.g. Allen \etal~\cite{allen93},
\cite{allen01a}) in terms of intrinsic absorption by  very cold gas
related to the strong cooling flow.  However our vision of
cool cluster cores has dramatically changed due to \xmm observations.
The standard cooling flow model predicts low energy emission lines
which are simply not seen in the RGS spectra (Peterson
\etal~\cite{peterson01}, \cite{peterson03}).  This standard model is
also inconsistent with EPIC data (e.g Molendi \&
Pizzolato~\cite{molendi01}; Matsushita \etal~\cite{matsushita02};
B\"ohringer \etal~\cite{bohringer02}; Kaastra et al.
\cite{kaastra03}).  No evidence of intrinsic
absorption was found with \xmm in the center of the cooling flow
regions in M87 and the Perseus cluster, and B\"ohringer \etal
(\cite{bohringer02}) argued that the excess absorption measured by
previous missions is an artifact of fitting standard cooling flow
models.  The low energy emission over-predicted by this model can be
artificially suppressed by adding an extra absorption component when
fitting spectra obtained with instruments like ASCA, which have
relatively low
sensitivity at low energies.

Although the absorption excess in A478 is confirmed by \xmm data, it
is more likely, in view of our current knowledge of cooling cores in
clusters, that all the absorption is of Galactic origin.  This
hypothesis is reinforced by the spatial distribution of the excess
absorption: the excess extends well beyond the cool core region.  
We also note that if cold gas has indeed now been detected in the core
of clusters, like A478, through CO measurements (Edge, 2001), there is
still a large mismatch, by an order of magnitude, between the inferred column
densities and the absorption excess (see Edge 2001, for full
discussion). Finally the local CXRB that we measure with \xmm is 
lower than the
average blank field value (see Sect.~\ref{sec:bkg} ).  The Rosat All
Sky Survey (RASS) $3/4~\keV$ maps (see Snowden~\cite{snowden97}) also
clearly show a deficit of CXRB in that region.  This again points
towards a high Galactic absorption.

       \begin{figure}[t]
         \hspace*{-1em}%
        \includegraphics[width=9cm]{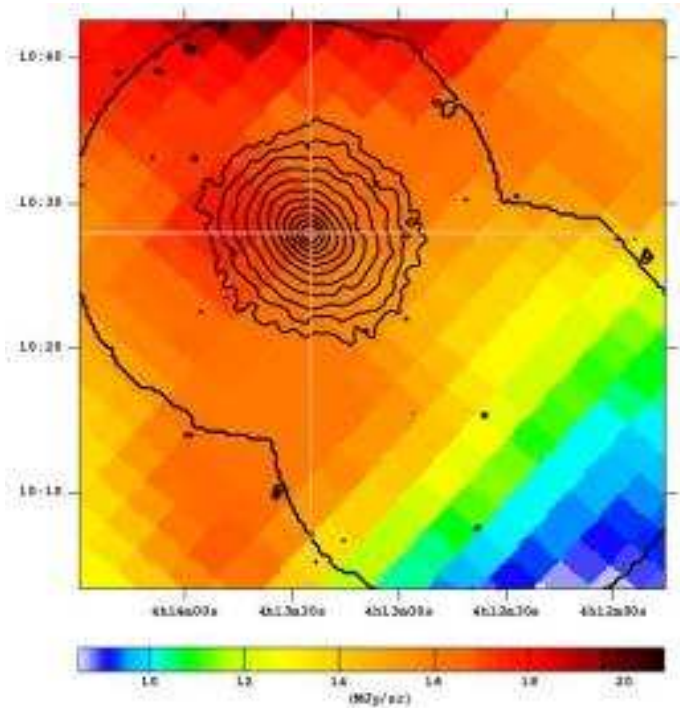}
        \caption{IRAS 100~$\mu$m (Schlegel et al. \cite{schlegel98}) in
          A478 direction.  The overplotted contours show the X-ray
          emission as well as the area covered by our \xmm
          observation. The vertical and horizontal lines defined the
          sectors used to cross-check the $\nh$ profiles (see text).
         \label{fig:f6}}
         \end{figure}

       \begin{figure}[t]
         \hspace*{-1em}%
        \includegraphics[width=9cm]{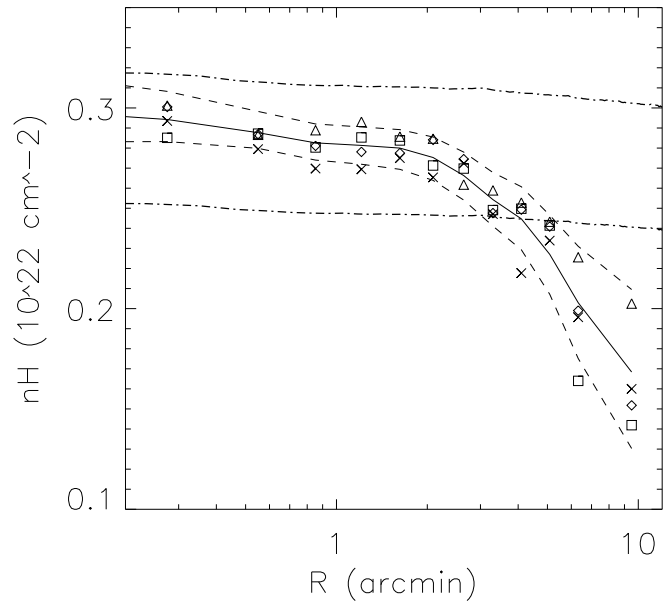}
        \caption{$\nh$ profiles derived from \xmm data in the North-East
          (diamonds), North-West (triangles), South-West (squares) and
          South-East (crosses) quadrant compared to the azimuthal
          profile (smoothed -- solid line).  The two dashed lines are
          the 3$\sigma$ error bars (smoothed) associated to the
          azimuthal profile.  The two dotted-dashed lines correspond
          to the $\nh$ value estimated from the 100~$\mu$m IRAS map
          according to Boulanger \etal (upper line --
          \cite{boulanger96}) and Schlegel \etal (lower line --
          \cite{schlegel98}).
         \label{fig:f7}}
         \end{figure}

To further test the origin of the absorption and the robustness of
our $\nh$ estimates, we considered other indicators of the Galactic gas
column
density.  We recall that the column density derived from X-ray data is
actually the total hydrogen column density (assuming standard
abundances).  We considered the IRAS 100~$\mu$m cleaned map of
Schlegel \etal (\cite{schlegel98}).  The Galactic hydrogen is
correlated with
the Galactic dust responsible for the IR emission, as shown by
Boulanger \etal (\cite{boulanger96}) and Schlegel \etal
(\cite{schlegel98}), who
correlated COBE/DIRBE and IRAS data with the Leiden/Dwingeloo survey
(Hartmann \& Burton~\cite{hartmann97}).  The correlation between the
IR emission and the atomic hydrogen column density is determined from
low $\nh$ data ($\nh < 4.6 \times 10^{20}$~cm$^{-2}$).  Above this
threshold an increasing dispersion is observed with higher IR/HI ratio
on average.  Boulanger \etal (\cite{boulanger96}) argued that this
excess IR
emission is due to dust associated with molecular Hydrogen. The IR emission
could thus actually be a tracer of the total hydrogen content.
Assuming that the correlation determined at low $\nh$ values (where the
H$_{2}$ fraction is expected to be small) is representative of this IR
- total $\nh$ correlation, we converted the IR brightness map into a
total $\nh$ map.  We used both the Boulanger \etal (\cite{boulanger96})
and
Schlegel \etal (\cite{schlegel98}) results, these two groups having
derived slightly different correlation coefficients.  We then derived
radial profiles which are compared with the \xmm derived $\nh$ profile
in Fig.~\ref{fig:f7}.  Interestingly, the IR and X-ray derived $\nh$
profiles are indeed found to be consistent up to about $5\arcmin$.

Beyond that radius the \xmm $\nh$ profile starts to become significantly
lower than the expected values from the IR emission.  However, the IR
emission shows a strong gradient over the cluster area in
the north-east/south-west direction (see Fig.~\ref{fig:f6}) .  There is
a drop by a factor of
two between the north-east sector of our mosaic and the south-east
sector.  Obviously the azimuthal average tends to smooth the gradient
effects.  Keeping the previous  definition of annuli, we then divided
each
annulus in four sectors separated by a North-South and an East-West
axis (see Fig.~\ref{fig:f6}). In each sector, we ran a spectral fit
for each annulus.  The
resulting $\nh$ profiles are presented in Fig.~\ref{fig:f7}.  All four
profiles are compatible with the azimuthally averaged profile within a
3$\sigma$ limit.  One can notice, however, that the South-East
measurements are systematically lower than the azimuthal values whereas
the North-West values are systematically larger.

In summary, both X-ray and IR data indicate a higher Galactic hydrogen
column density than the 21 cm value.  Both X-ray and IR $\nh$
estimates agree remarkably well in the cluster center, suggesting that
the excess absorption is indeed of Galactic origin.  Moreover, FIR
observations from the ISO satellite at 90 and 180~$\mu$m show a color
ratio favoring a cold temperature structure (Pointecouteau \& Giard,
in preparation), which is more likely to be due to a Galactic
structure than to an intracluster dust component.  Indeed, the
expected temperatures for the intracluster dust according to the current
models are  $>20$~K (\cite{montier03}).  However, the radial
variations of the X-ray and IR derived hydrogen column density differ.
This could be due to variations in gas to dust ratio and/or
metallicity for instance.
To probe the foreground structure on the cluster scale toward A478,
FIR observations with a higher spatial resolution would be extremely
useful.  For instance, soon the ASTRO-F mission (\cite{shibai02}) will
survey the whole sky at FIR wavelengths and will provide observations
up to 200$~\mu$m that will reveal the galactic cold component
structure.  These upcoming observations on the whole cluster scale will
certainly help to clarify this issue.

In view of the discrepancy between the IR and X-ray derived $\nh$
values beyond $\theta \sim 5\arcmin$, we further checked the
robustness of our $\nh$ and thus $\kT$ measurements in that region
(two outer annuli).  We ran again the individual spectral fit on the
annuli, fixing the value of the Galactic absorption to $2.5~\times
10^{21}$~cm$^{-2}$.  The derived temperature drops from $\sim 7~\keV$
down to $\sim 5~\keV$. However, the fit is significantly worse.
Indeed, the F-test probability, given the $\chi^{2}$ value and the one
obtained previously by letting free the $\nh$, is 10$^{-8}$.
Furthermore, if we still fix the  $\nh$ value to $2.5~\times
10^{21}$~cm$^{-2}$,  but fit the spectra for
$E>1.5$~keV, avoiding the low energy band that is sensitive to the
Galactic absorption, the derived temperature profile is fully
compatible with the nominal one.  Therefore, we can be reasonably
confident that the derived radial column density and temperature
profiles are real.

\subsection{Comparison with previous results}

The best fit values we derived for the overall physical properties of
A478 are in agreement with the previous results for the
Galactic absorption and the metallicity (see Table~\ref{tab1}).
Moreover, their radial profiles match
closely those derived by Sun et al. (\cite{sun03}) using the
Chandra data. However, our overall temperature value is
marginally compatible with the temperature values from ROSAT/Ginga
    and ASCA.
A478 does not exhibit an overall isothermal plasma (presence of a cool
core).  The temperature derived from an isothermal fit is actually an
emission weighted temperature, which depends on the instrument response.
Due to its higher sensitivity at low energies, XMM is more sensitive to
the presence of a cool component. This would  explain the slightly
lower temperature derived from our data  with respect to Ginga and ASCA
results.
To further check this point,  we fixed the absorption value to
$2.5\times 10^{21}$~cm$^{-2}$ (the value derived from the ROSAT/Ginga
analysis) and we fitted the overall spectrum over the [1-10]~keV energy
band. The best fit temperature value is then  $kT=6.42\pm 0.06~\keV$, a
value compatible with ROSAT/Ginga value, as well as ASCA
value.

Despite this agreement, some important discrepancies  appear between
the temperature
profiles obtained from \xmm and Chandra. If we focus on the average
value of the temperature
excluding the cool core region, the Chandra value ($\sim8.5~\keV$)
is  significantly higher than the \xmm value ($\sim6.5~\keV$).
Similar discrepancies appear for the luminous X-ray cluster
PKS 0745-191. Indeed, the temperature derived from the
Chandra analysis (\cite{hicks02}) outside the cool core
($1.5'<r<2.3'$), $\sim10.5~\keV$, is significantly higher than the value
derived from \xmm data (\cite{chen03}) in the same
region, $\sim7.5~\keV$. In this case, the \xmm result compares better
with the
value of $\sim8~\keV$ by BeppoSAX (\cite{degrandi99}). However, it must
be noted that
the BeppoSAX temperature measurement includes the
cold core region which is likely to induce a bias toward lower
temperatures.

We have failed to explain the discrepancy between \xmm and Chandra.
Apart from calibration related problems, we thought that it could be
due to background subtraction problems.  For both PKS 0745-191 and
A478, the CXRB was found with \xmm ~to be different from the CXRB of
a typical blank field.  The higher $\nh$ observed in the direction of the
A478 cluster certainly contributes to this difference for this
cluster, but we cannot exclude a contribution from some intrinsic
spatial variation in the soft X-ray Galactic emission.  This
difference was taken into account in the \xmm background subtraction
procedure (in the second subtraction step -- see Sect.~\ref{sec:bkg}).
On the other hand, Chandra analysis had to rely on a simple blank
field background subtraction, by lack of data at large radii.  In the 
Chandra analysis of A478, the CRXB is a priori oversubtracted and this could
bias the temperature determination, especially in the outer cluster
region where the CXRB count rate is no longer negligible with respect to
the cluster count rate.  To test this hypothesis for A478, we perform
a single blank field subtraction for each XMM annular spectrum and
re-ran the spectral isothermal fit.  The resulting absorption and
temperature profiles become significantly different beyond $5\arcmin$
(the upper limit of the Chandra profile).  However, below
$\sim5\arcmin$, the profiles are not significantly affected and remain
inconsistent with Chandra values\footnote{Our background subtraction
  procedure assumes that the CRXB does not vary within the field of
  view.  The low value of the local CRXB (as compared to blank field
  value) may partly be due to a higher galactic absorption
  (Sect.~\ref{sec:nh}).  We thus cannot exclude that the CRXB actually
  varies within the FOV, in view of the observed $\nh$ variations (see
  Fig.~\ref{fig:f7}).  However, this further test shows that the
  double subtraction is really needed only for the cluster outskirts
  ($5\arcmin-12\arcmin$), closest to the region chosen to compute the
  local background ($r>16$').  This should minimise any artifact due
  to our neglect of possible background radial variation.}.  Although
the Chandra blank field observations are not the same as those of {\it XMM-Newton},
it is thus unlikely that the background subtraction issue is an
explanation of the discrepancies.  Furthermore the agreement between
the \xmm and Chandra profiles for the Galactic absorption would be
puzzling if that was the explanation, since the $\nh$ determination is
particularly sensitive to the subtraction of the residual CXRB, which
affects the low energy part of the spectrum most.

\subsection{The cluster temperature profile: correction of PSF and
projection effects. \label{sec:specor}}

        \begin{figure*}[t]
\centering
         \hspace*{-2em}%
%  \epsfxsize=17.5cm
%   \epsfbox{FigkTprof_PSFProj.eps}
\includegraphics[height=9.5cm]{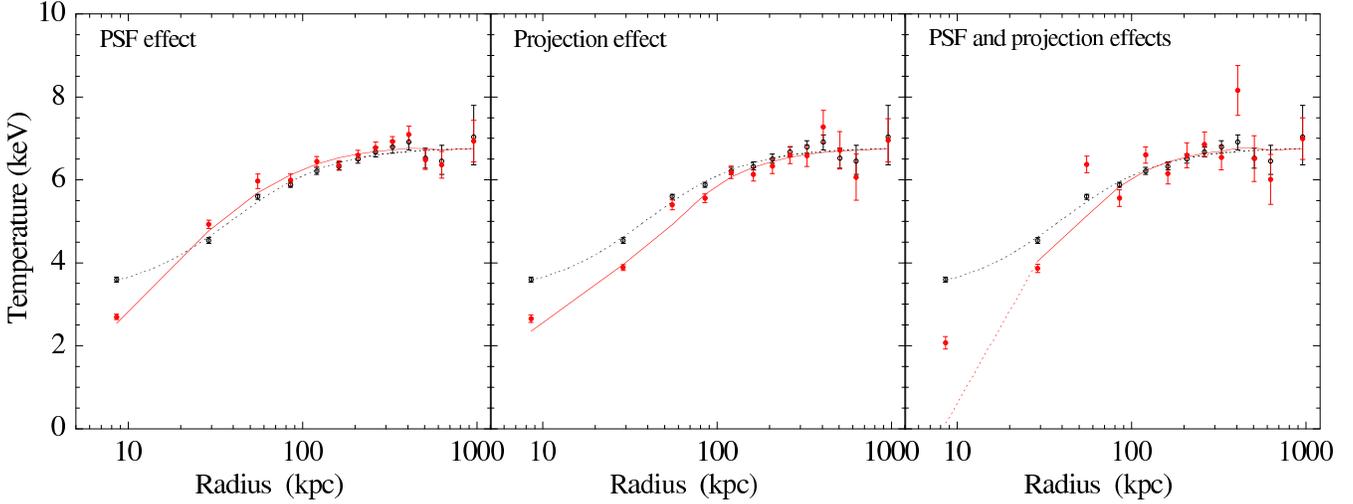}
\vspace*{-9em}
        \caption{Temperature profiles corrected for the PSF effect (left
          panel), deprojected (middle panel), PSF corrected and
          deprojected (right panel).  The data points are the
          temperatures derived from simultaneous fitting of the
          annular spectra.  The open points are the temperatures
          obtained by fitting each annular spectrum individually with
          an isothermal model (Same as in Fig.~\ref{fig:f5}).  The
          dotted line is the analytical model fitted to these data
          (Eq.\ref{eq:fitkt}).  The full lines are the PSF corrected model
          (left panel), the deprojected model (middle panel) and the
          PSF corrected deprojected model (see Sect.~\ref{sec:modcor}
          for details).
         \label{fig:f8}}
         \end{figure*}

The central drop in the temperature profile emphasizes the
need to perform a deprojection analysis and to take into account the
PSF effect.  For such a highly peaked cooling flow cluster, PSF
effects are important in the center where we have chosen narrow bins
for the
temperature computation to recover the best temperature profile.

\subsubsection{PSF and projection effect modeling}

The incident emission of each annulus, $i$, is the projected sum of
the emission from various shells, $j \geq i$.  This emission is then
redistributed among various annuli due to the finite PSF. In principle,
we should apply an absorption model to each incident annular spectrum
(i.e.
after the projection and before the PSF convolution).  However the
$\nh$ profile derived from the annular spectra fit does not show strong
gradients in the central region (where the PSF effect is significant).
We thus used a single absorption model for each {\it observed} annular
spectrum, fixing the $\nh$ value to the best fit value obtained from
individual annular fits (see Table~\ref{tab2}).

We thus model the observed annular spectra, $S_{\rm i}^{\rm O}(E)$,
with a linear combination of isothermal \mekal\ models (normalized to
the unit emission measure), multiplied by a \wabs\ model:
\begin{equation}
S_{\rm i}^{\rm O}(E) = \wabs(\nh_{i}) \sum_{j=1}^{n} a_{i,j} \mekal
(T_{j}, Z_{j})
\label{eq:multifit}
\end{equation}

We first considered pure PSF effects.  In that case, each $a_{i,j}$
coefficient is the emission measure contribution of the ring $j$ to
the ring $i$.  The fitted temperatures $T_{j}$ can be considered as
`PSF corrected' projected temperatures.  The $a_{i,j}$ factors were
derived from our best fit gas density profile (BBB model), converted
to an emission measure profile and convolved with the \xmm PSF
estimated at $1~\keV$.  To crudely validate our absorption modeling,
we ran this PSF correction fit leaving the $\nh$ as a free parameter.
We found completely consistent values with the best fit value of the
annular spectral fit.

Similarly we assess the projection effects, neglecting the PSF
blurring.  We used the same formula with the $a_{i,j}$ redistribution
factors being now the emission measure contribution of the shell $j$
to the ring $i$ and $T_{j}$ the temperature of the shell $j$ (assumed
to be isothermal).

Finally we took into account both effects, using as
$a_{i,j}$ the emission measure contribution of the shell $j$ to the
ring $i$ after convolution with the PSF.

\subsubsection{Simultaneous fits of annular spectra}

The fitting was done with \xspec.  We have to take into account that
\xspec~ can only handle 1000 parameters (even if most of them are
frozen).  The EMOS1, EMOS2 and EPN spectrum of each annulus was loaded
into \xspec~ as a data group.  The same model parameters are applied to
each spectrum of a given group.  Therefore, EMOS and EPN spectra have
to be normalized in order to be fitted with the same normalization.
Furthermore the $a_{i,j}$ coefficients are computed without taking into
account flux loss due CCD gaps, bright pixels, etc\ldots.  We therefore first
renormalized each spectrum by the ratio of the annular geometrical
area to the actual extraction region surface ({\sc backscale} value).
After this correction, the ratio of the EMOS and EPN normalizations
obtained from the annular fits (Sect.~\ref{sec:speraw}) were found to be
consistent with the ratio obtained by fitting the overall EMOS and EPN
spectra: $N_{EMOS/EPN} = 1.11$.  We thus applied this factor to all
EPN spectra.  Finally, we checked that the annular fit results
indeed remain the same: the differences in derived \kT\  are negligible
compared to the statistical errors.

Fitting simultaneously $n$ annular spectra with a sum of $n$ \mekal\ 
models, multiplied by a \wabs\ model, gives a total number of
parameters of $(6n + 1)n$.  We thus have to limit $n$ to 12.  To
overcome this problem, we have used two different sets of 12 annular
spectra.  The first set is obtained by grouping the last two annuli
into a single annulus and the second set by grouping the first two
annuli.  We then combined the first set results for annuli $\#1$ to
$\#9$ with the second set results for annuli $\#10$ to $\#13$.

For each set, the free parameters are the 12 temperatures, 12
normalizations, one per data group (annulus), the other normalizations
being linked according to Eq.~(\ref{eq:multifit}).  In practice we
ignore all contributions less than $1\%$.  We have frozen the
abundance of each \mekal\ model to the annular best fit value, except
for the 4 innermost components corresponding to the $r <1'$ region.
Beyond that region the annular abundance profile is flat.  PSF and
projections effects are unlikely to affect the abundance determination
in such a way that it has a significant impact on the temperature
determination.  To further check this point, we also fixed all the
abundances to the best fit annular value.  Only the central
temperature is changed significantly.

The resulting corrected temperatures are plotted versus effective
radius (defined in Eq.~\ref{eq:rw})  in Fig.~\ref{fig:f8}.  These data are
compared to the temperature profile derived in Sect.~\ref{sec:speraw}
(hereafter the raw temperature profile).

\subsubsection{Correction of the temperature profile model}
\label{sec:modcor}

We also considered the best fit model
(Eq.~\ref{eq:fitkt}) of the raw temperature profile (dotted line in
Fig.~\ref{fig:f8}).  It can also be corrected for PSF and/or
projection effects, assuming that the annular temperatures $T_{\rm
i}^{\rm M}$ are emission-weighted temperatures:
\begin{equation}
T_{\rm i}^{\rm M} = \sum_{j=1}^{n} \frac{a_{i,j}}{\sum_{j=1}^{n}
a_{i,j}}
T_{j}
\label{eq:multikT}
\end{equation}
The factors $a_{i,j}$ are the same as computed above.  The
PSF-corrected model, the deprojected model and the PSF-corrected
deprojected
model are plotted in Fig.~\ref{fig:f8} (full lines).\\

\subsubsection{Results \label{sec:speres}}

First, it is instructive to consider the PSF correction
(Fig.~\ref{fig:f8}, left panel) and the deprojection
(Fig.~\ref{fig:f8}, middle panel) separately.  In both cases, the
corrected
model fits reasonably well the corresponding corrected temperature
profile derived from the spectral fit.  However, while the former
remains a smooth function of radius, the later is more noisy.  A
comparison of the observed and model profiles before and after
correction shows that the correction process amplifies any variation
of the raw temperature profile compared to the smooth model (see in
particular the temperatures of annuli \# 3, \#10 \#11 and \#12).  This
is probably linked to the well known problem of noise amplification when
deconvolving or deprojecting  noisy data (see also Kaastra et al.
(\cite{kaastra03}) for a discussion on PSF/projection effects).

The PSF-corrected projected temperatures and the deprojected
temperatures are  consistent, within the error bars, with the raw
temperatures in the external part of the cluster (nearly isothermal
region, $R > 100$~kpc).  Significant deviations are seen for the
first four annuli.  The general effect of the PSF blurring and of the
projection are best seen by comparing the corrected and uncorrected
models.  The PSF affects mostly the central bin (the temperature of
which is increased due to contamination by the higher temperatures
of the external bins).  In turn the other bins are contaminated by the low
temperature central bin, and their observed raw temperatures are
slightly lower than the incident ones.  The effect is small however
and consistently tends to zero with increasing radius.  The main
effect is the projection effect.  As expected it damps down the
gradient in the central region.  The deprojected temperatures are always
smaller than the projected ones, the effect increasing with decreasing
radius.

The PSF-corrected deprojected temperature profile is shown in
Fig.~\ref{fig:f8} (right panel).  When both PSF and projection effects
are taken into account the noise amplification is dramatic.  The
temperature profile derived from the spectral fit shows strong
discontinuities (e.g. bins \#3,4,5, bin \#10).  On the other hand, the
corrected model remains smooth.  Although the corrected temperatures
derived from the spectral fit are not consistent with the corrected
model within the statistical errors, they are distributed around
it.  The largest deviations correspond to the discontinuities
mentioned above and are again clearly located around the bins which
originally deviate most from the smooth model profile.  These
discontinuities are thus very likely non-physical and the corrected
model is probably a better representation of the true cluster
temperature
profile than the profile derived from the spectral fit. As
discussed in Sect.~\ref{sec:mass}, we will use this corrected model as a
reference in our computation of the mass profile.
~\\
\indent The PSF-corrected deprojected model profile is consistently intermediate
between the PSF-corrected model profile and the deprojected model
profile for bins $\#2$ to $\#13$.  As shown above, the PSF and
projection have opposite effects in that region.  The model temperature
of the central bin is however extremely low.  The PSF correction and
deprojection was done assuming that the annular temperatures are
emission-weighted temperatures.  This assumption is probably less and
less valid with decreasing temperature and the low value we derive
might be an artifact of our assumption. Furthermore the gas may be
multiphase in that region ($r<20$~kpc) due to the interaction of the
intracluster medium with the central cD galaxy  (see also below).

%__________________________________________________________________

\section{Mass profile \label{sec:mass}}

\subsection{Calculation of the total mass profile}
\begin{figure}
\begin{centering}
\centering
\hspace{-2em}%
\includegraphics[width=9cm]{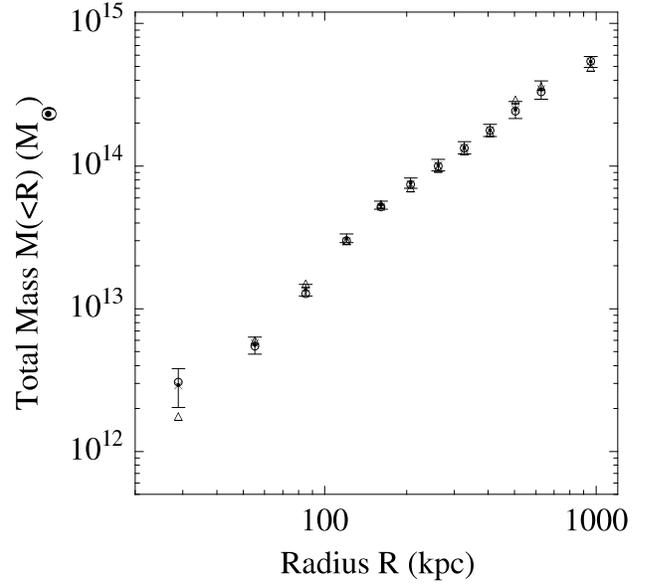}
\vspace{-9em} \caption{{\footnotesize Integrated total mass
distribution.  Filled circle: The reference total mass profile
obtained from the best fit BBB model for the gas density profile and
the PSF-corrected deprojected model of the temperature profile (errors
are $1\sigma$).  Open circles: mass obtained from the deprojected
temperature model profile; Triangles: mass obtained using the
deprojected temperature profile derived from the spectral fit.
Crosses: mass using the Chandra gas density profile.  }}\label{fig:f9}
\end{centering}
\end{figure}

The total gravitational mass distribution shown in
Fig.~\ref{fig:f9} was calculated under the usual assumptions of
hydrostatic equilibrium and spherical symmetry using

\begin{equation}
M(r) = - \frac{kT\ r}{{\rm G} \mu m_p}   \left[ \frac{d \ln{n_{\rm
g}}}{d \ln{r}} + \frac{d \ln{T}}{d \ln{r}} \right]
\label{eq:HE}
\end{equation}
\noindent where G and $m_p$ are the gravitational constant and proton
mass, and $\mu = 0.609$.  The mass (with errors) at each radius was
calculated with the Monte Carlo method described in Pratt \&
Arnaud~(\cite{pratt03}), which takes as input a parametric model for
the gas density profile and a measured temperature profile with error
bars.  A random temperature at each radius of the measured temperature
profile is generated assuming a Gaussian distribution with sigma equal
to the $1\sigma$ error and a cubic spline interpolation (between 3
adjacent points) is used to
compute the derivative.  Only `physical' temperature profiles are
kept, i.e. those yielding  monotically increasing total mass
profiles.  In total 5000 such profiles were calculated.

At large radii, the errors on the derived mass profile are dominated by
the statistical
errors  on the temperature profile.  However
we have to consider possible systematical errors, specially  in the
central
($R<100$~kpc) region, where the PSF and deprojection corrections
introduce noise in the derived temperature profile (see Sect.
\ref{sec:specor}).  Only data beyond $30~{\rm kpc}$ are considered:
the temperature of the central bins is highly uncertain (see Sect.
\ref{sec:speres}).  Furthermore, Chandra data have clearly revealed
sub-structure in the central $\theta \sim 20~{\rm kpc} = 0.2\arcmin$
region.  Below that radius, the hot thermal gas might interact with the
radio halos, producing a non-thermal population of electrons, and
therefore the hydrostatic equilibrium might be disturbed locally
producing possible multiphase states for the gas at this spatial scale
(see previous work on M87: \cite{bohringer95,belsole01} and
PKS~0745-191: \cite{chen03}).
%% Figure: total,gas mass
%%

\begin{figure}
\begin{centering}
\centering
\hspace{-2em}%
\includegraphics[width=9cm]{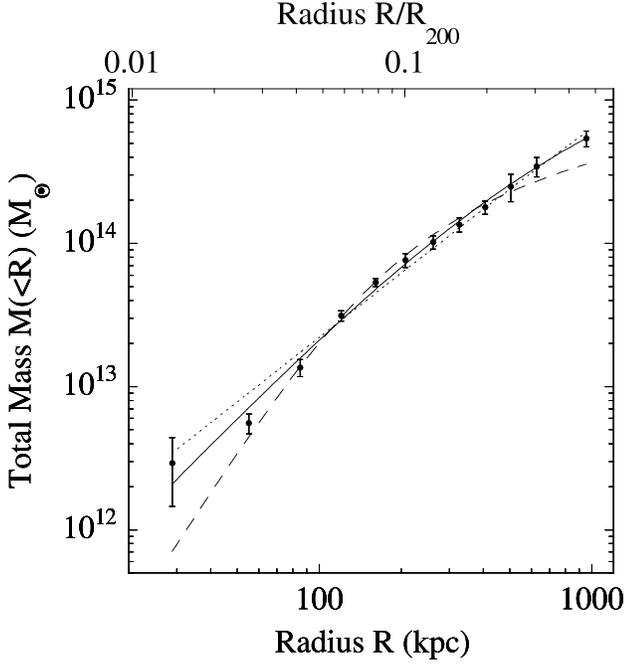}
\vspace{-9em}
\caption{{\footnotesize
Modelling of the integrated total mass distribution.  Filled circle:
total mass profile as in Fig.~\ref{fig:f9}.  The errors now include
systematic errors due to the PSF correction (see text).  The solid line
is the
best fit NFW profile ($c=4.18$), the dotted line is the best MQGSL
profile and the dashed line is the best fit King model.
}}\label{fig:f10}
\end{centering}
\end{figure}
%%
%%===========
%%

We first derived a reference mass profile.  It was computed using our
best fit BBB model for the gas density profile.  For the temperature
profile we used the PSF-corrected deprojected model profile. The
profile derived from the spectral fit is too noisy to be used: the
strong fluctuations observed are inconsistent with any underlying mass
profile.  It is difficult to assess the statistical errors on this
model profile in an objective way.  We used the errors derived from
the spectral fit, which is probably a conservative approach.  For bin
$\#2$ we also add as an error the difference between the fitted
temperature and the model value (the latter might be affected by the
too low value derived for bin \#1).

We then computed the mass profiles obtained using alternative
temperature and density data, to assess the systematic uncertainties.
For the gas density profile, we considered the best fit BBB model of
the Chandra profile (crosses in Fig.\ref{fig:f9}). The differences
between the derived mass profile and the reference profile are much
smaller than the statistical uncertainties.

We then considered systematic uncertainties due to the temperature
profile determination. We have shown that PSF effects in the region
considered here are less important than projection effects and that
pure deprojection introduces much less noise.  We thus also computed
the mass profile derived from the deprojected temperature profile
(neglecting the PSF blurring).  We used both the profile derived from
the spectral fit (triangles in Fig.\ref{fig:f9}) and the deprojected
model profile (open circles in Fig.\ref{fig:f9}).  The latter is well
within the error bars of the reference profile, but the former differs
significantly (see for instance the first point).  Considering
that these differences are likely to be representative of the
systematic uncertainties due to the PSF correction treatment, we add
them quadratically to the statistical errors on the reference mass
profile.

In the following, we will thus consider the reference mass profile
with these errors bars.

\subsection{Modeling of the total mass profile}
We first tried to fit the data with a King model, where the mass
density profile is  given by: $\rho(r) \propto \left[1+
(r/\rs)^2\right]^{-3/2}$.   This model (dashed line in
Fig.~\ref{fig:f10}) is inconsistent with our data, the $\chi^2$ is
$\chi^2 = 16.5$ for $10$ d.o.f.

The total mass profile was then fitted using cusped density
distributions: the Navarro, Frenk \& White (\cite{navarro97}) profile
($\rho(r) \propto \left[(r/\rs) (1+ r/\rs)^2\right]^{-1}$) and the
Moore \etal (\cite{moore99}) profile ($\rho(r) \propto
\left[(r/\rs)^{3/2} (1+ r/\rs)^{3/2}\right]^{-1}$).  These models have
two free parameters, the central density and the scaling radius, or
equivalently the total mass $M_{200}$ (corresponding to a density
contrast of 200, as compared to the critical density of the Universe
at the cluster redshift) and the concentration parameter
$c=R_{200}/\rs$.  All useful formulae relating these quantities can
be found in Pratt \& Arnaud (\cite{pratt02}).

The NFW profile provides a good fit to the data: $\chi^2 = 9.5$ for
$10$ d.o.f.  The best fit NFW parameters are: $r_s=492\pm71~{\rm kpc}$
and $R_{200} = 2076\pm106~{\rm kpc}$, corresponding to a concentration
parameter of $c = 4.22 \pm 0.4$ and a total mass enclosed within
$R_{200}$ of $M_{200} = 1.1 \times 10^{15}$~M$_{\odot}$ .  The
previous errors are quoted at 68\% confidence level.  This best-fit
NFW model is shown overplotted on the mass profile of the cluster in
Fig.~\ref{fig:f9}.  The upper axis is the radius in units of the
derived $R_{200}$.  We are thus probing the dark matter shape on a
scale from $\sim 0.01$ to $\sim 0.5$ virial radius.

The alternative MQGSL profile is rejected by our data: $\chi^2 =
29.7/10$.  It must also be noted that the deviations from the data are
not only significant at low radii, where the mass estimate is most
sensitive to systematic errors.  There is a general deviation at large
radius, where the model gives essentially a power law, while the
observed profile shows a significant curvature.

%% Figure: gas mass fraction
%%
\begin{figure}
\begin{centering}
\centering
\hspace{-2em}%
\includegraphics[width=9cm]{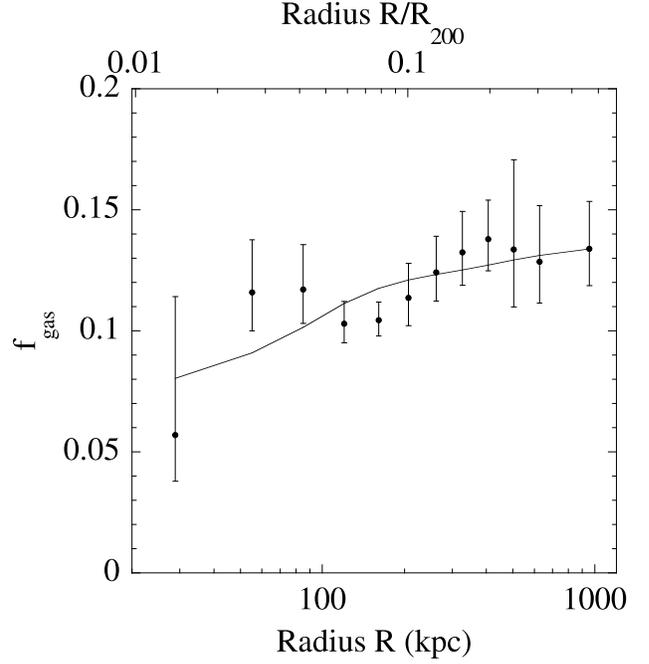}
\vspace{-8.5em}
\caption{{\footnotesize
    The integrated gas mass fraction as a function of the radius. The
    1$\sigma$ error bars are obtained from the propagation of the
    errors over the total mass profile.  The gas mass fraction profile
    computed from the best fit NFW model is plotted as a solid line.
  }}\label{fig:f11}
\end{centering}
\end{figure}
%%
%%===========
%%

\subsection{Gas mass fraction profile}

We derived the integrated gas mass fraction profile, $\fgas$ , from
the ratio of the gas mass profile to the total mass profile. The gas
mass profile is computed from the integration of the best fit BBB
model for the density profile (see Sect.~\ref{sec:nefit}). The errors
for the gas mass fraction are obtained from the propagation of the
total mass uncertainties, the gas mass uncertainties being negligible.
The resulting gas mass fraction profile is shown in
Fig.~\ref{fig:f11}, together with the profile derived from the NFW
best fit model.  There is a general increase of $\fgas$ with radius,
although the effect is small over the [0.01-0.5]~$R_{200}$ range. The
gradient is most pronounced in the cool core region (about $30\%$),
beyond which there is only marginal evidence of a positive gradient
($\sim 10\%$ from $0.1$ to $0.5~R_{200}$).

Excluding the cool core region (e.g. $r<2$\arcmin), the average value
for the gas fraction is $\bar{f}_{gas}(r>2{\rm
  \arcmin})=0.129\pm0.008$.  We interpolated the gas mass fraction
value at $r_{2500}=822~{\rm kpc}$, the radius corresponding to a
density contrast of $\delta=2500$ (2500 times the critical density of
the Universe at the cluster redshift), $f_{2500}=0.13\pm0.02$.  This
value agrees with the average value of $0.113\pm0.013$ at the same
overdensity derived by Allen \etal (\cite{allen02b}) from a sample of
six massive clusters observed by Chandra.  Our value is also
consistent with the value derived from the analysis of A1413 (Pratt \&
Arnaud \cite{pratt02}) when scaled to the chosen cosmology of this
paper: $f_{2500}(A1413)=0.11$.  However, the A478 and A1413 $\fgas$
profiles have different shapes, the latter decreasing more
strongly toward the center. This reflects differences in the
central gas density distribution, the total mass profile of both
clusters having similar concentration parameters.  Such a difference
is probably linked to the different thermodynamical states of A1413
and A478, the latter hosting a strong cooling core, contrary to the
former.

%__________________________________________________________________

\section{Discussion and conclusion}

In this paper we have analysed imaging and spatially resolved
spectral data of the galaxy cluster A478 obtained with the \xmm
satellite.  We obtained well constrained absorption, gas density and
gas temperature profiles up to $\sim 0.5$ virial radius.

As in previous studies, we found an excess of absorption in the
direction of A478.  The derived absorbing column density exceeds the
21~cm measurements by a factor of $\sim2$ in the center and the excess
extends well beyond the cooling core region.  This excess of
absorption seen in A478 (and other cooling flow clusters) was
interpreted in pre-XMM and Chandra studies (e.g. Allen \etal
\cite{allen93,fabian94}) as the signature of intrinsic cool absorbing
material, a consequence of the strong cooling flow in the cluster
center.  From the absorption excess extent and a detailed comparison
with FIR data, we argue that the absorption excess is rather of
Galactic origin. We suggest it could be the effect of a Galactic
molecular/cold cloud type structure in the line of sight.  The next
generation of FIR space missions will help to clarify this issue with
sensitive FIR mapping of the whole cluster area with a high spatial
resolution.

We fitted the surface brightness profile with various parametric
models of the gas density profile, taking into account  the \xmm PSF.
    The gas density profile, derived on scales of  $0.03\arcmin -
13\arcmin$,   is highly peaked towards the center and is well fitted by
a quadratic sum of three \betamod.  The derived gas density profile is
in excellent agreement both in shape and normalization with the
Chandra density profile (measured up to  $5\arcmin$ of the center).
This indicates that the PSF modeling we have used is basically correct
and that accurate density profiles in the very center of the cluster can be
derived with XMM data, in spite of the PSF blurring.

A raw temperature profile was obtained on scales of  $0.07\arcmin -
10\arcmin$  by fitting isothermal models to spectra extracted in $13$
concentric annuli. This profile  shows  a sharp negative gradient
measured toward the center ($r<2\arcmin$), a  signature of a cooling core.
Beyond that region the profile is essentially flat. We have thoroughly
investigated projection and PSF effects on the temperature profile
determination. The PSF effects beyond $0.3\arcmin$ are much less
important than projection effects, whereas both are important in the
very center. We discuss the noise introduced by the correction of these
effects and a way to overcome this problem.    The derived deprojected
PSF-corrected temperature profile ranges  from  $\sim2$~keV in the
center up to an asymptotic value of $\sim6.5$~keV.

Using this temperature profile and the density profile, we have derived
the total mass profile
for this cluster from $0.01$ up to $\sim0.5$ times the virial radius.
Systematic uncertainties due to the PSF and projection correction for
the temperature profile are taken into account. We have tested
different models for the dark matter profile distribution against the
observed mass profile. A mass distribution with a cusp in the center,
as predicted from numerical simulations, is clearly preferred.

An isothermal sphere model does not provide a good fit to the data.
In a second time we tested an MQGSL model and an NFW model.  Those two
types of models have similar shapes at large radii (they both scale
like $r^{-1.5}$) but differ significantly at small radii. Therefore to
discriminate those two models one needs data with a high statistic
quality over a wide range of radii (i.e. covering at least two
decades).  Our data set nearly fulfilled this requirement and we were
indeed able to discriminate between the two models, the NFW model
being preferred to the MQGSL model. For the NFW model, we derived a
concentration parameter $c=4.2\pm 0.4$. This value is as expected from
numerical simulations: $c \sim6$ (Navarro, Frenk \& White
\cite{navarro97}; \cite{eke98}) with a typical $1\sigma$ dispersion of
$\Delta(\log(c))=0.18$ (Bullock \etal~\cite{bullock} ). This work can
be compared with the similar work on the cluster A1413 by Pratt \&
Arnaud (\cite{pratt02}). In the case of A1413, if the NFW model was
acceptable, the MQGSL model was slightly preferred. Although this
cluster is detected out to 0.7 times the virial radius, the data are
limited in the center, a shortcoming, as emphasized above, for
discriminating between those two models. Moreover, data in the center
only are not sufficient (see the work on A1983 by Pratt \& Arnaud
\cite{pratt03}).  On the other hand, our result agrees with the
analysis of A2029 by Lewis et al. (\cite{lewis03}), which clearly
favors an NFW dark matter profile. To our knowledge, this is the only
other data set which covers a similar wide radial range (0.001 to 0.1
virial radius) .

The key factors in fitting the mass profiles with different dark matter models,
%to fully probe the different dark matter models against the
%shape of the dark matter profiles, 
are the resolution in the center as
well as the data at large radii.
To date \xmm  is the best satellite to compute total mass
profiles, especially through its capability to derive precise
temperature profiles. Nevertheless, its spatial resolution limits the
investigation at the very center of galaxy clusters. A direct combined
analysis of \xmm and Chandra data of very well relaxed clusters seems
to be an ideal path to a full description of the dark matter profile in
clusters. However,  one has to keep in mind that this requires an
excellent cross calibration between the two satellites, so that the
temperature profiles derived at various scales can be combined.

%__________________________________________________________________

\begin{acknowledgements}
The authors thank  A.~C. Edge, the referee, for his useful comments 
ans suggestions.
We thank M. Sun for providing the Chandra density profile and useful
discussion. EP and MA are grateful to G.~W. Pratt for his help and
fruitful discussions throughout this work. We thank A. Sanderson for useful
discussion.
EP acknowledges the support of CNES, the French Space Agency.
This work is based on observations obtained with \xmm, an ESA science
mission with instruments and contributions directly funded by ESA
member states
and the USA (NASA). The Space Research Organization of the Netherlands
(SRON) is supported financially by NWO, the Netherlands Organization for
Scientific Research.

\end{acknowledgements}

      \appendix

\section{Extraction of vignetting-corrected products from merged event
lists \label{sec:app}}

To correct for vignetting effects we used the photon weighting method,
described in Arnaud \etal~(\cite{arnaud01}).  An estimate of the
vignetting-corrected number of counts in a given sky region $Reg$ and
in a given energy band $E_{min}-E_{max}$ is the sum:
\begin{equation} C = \sum_{j} w_{j}
\label{eq:w}
\end{equation}
over all events $j$ with sky position $(x_{j},y_{j}) \in Reg$ and
energy $E_{j} \in [E_{min}-E_{max}]$.  The weight coefficient $w_{j}$
is the ratio of effective area at the event position to the central
effective area computed at the event energy.  This count extraction is
the same for individual or merged events lists.  We computed the
weight
coefficients by running the task \evigweight\ on each individual events
list
(this task can also be run on the merged events lists).

Count rate estimates are less straightforward, since the effective
exposure time can strongly vary in the extraction region (from 11~ksec
in regions only observed with the offset pointings up to 60~ksec in
the overlapping region).  However, the total count rate in a given
region is simply the sum of the count rates in various sub-regions.
The count rate can be written as
\begin{equation} CR = \sum_{j}
\frac{w_{j}}{t((x_{j},y_{j})}
\label{eq:corspec}
\end{equation}
where $t(x_{j},y_{j})$ is the exposure time at the event location.

In practice we used the following convenient procedure for each
camera data set:

\noindent 1 - We created a mosaic exposure map of the two pointings
in sky
coordinates.  The reference position is the same as for the merged
events list.  The exposure map takes into account detector regions
excluded by the events selection criteria (bad pixels, CCD borders
\ldots ).  We used a pixel size of $1.1"\times1.1"$

\noindent 2 - After merging the events lists, we divided the weight
coefficient of each event by the exposure time taken from this
exposure map.

\noindent 3 - Scientific products (spectra, images, profiles) in count
rates can be readily extracted from the merged events list by binning
the events
weighted by these new weight coefficients.  As these products are
corrected for vignetting, we can then used the on-axis response for
further
physical analysis.


\begin{thebibliography}{}

\bibitem[1993]{allen93} Allen , S.~W., Fabian, A.~C., Johnstone,
R.~M., \etal\ 1993,
\mnras, 262, 901

      \bibitem[2001a]{allen01a} Allen, S.~W., Fabian, A.~C., Johnstone,
R.~M., \etal\ 2001, \mnras, 322, 589

\bibitem[2001b]{allen01b} Allen, S.~W., Schmidt, R.~W., \& Fabian,
A.~C.\ 2001, \mnras, 328, L37

\bibitem[2002]{allen02a} Allen S.W., Schmidt, R.W., Fabian, A.C., 2002,
               \mnras, 335, 256

\bibitem[2002]{allen02b} Allen, S.~W., Schmidt, R.~W., \& Fabian,
A.~C.\ 2002, \mnras, 334, L11

\bibitem[2002]{abg02}
               Arabadjis, J.S., Bautz, M.W., Garmire, G.P., 2002, ApJ,
572, 66

\bibitem[2001]{arnaud01} Arnaud, M., Neumann, D.~M., Aghanim, N.,
\etal\ 2001, \aap, 365, L80

\bibitem[2002]{arnaud02b} Arnaud, M., Majerowicz, S., Lumb, D. \etal\
2002,
\aap, 390, 27

\bibitem[Belsole et al. (2001)]{belsole01} Belsole, E., \etal\ 2001,
\aap, 365, L188

\bibitem[B{\"o}hringer et al. (1995)]{bohringer95}
B{\"o}hringer, H., Nulsen, P.~E.~J., Braun, R., \& Fabian, A.~C.\ 1995,
\mnras,
274, L67

\bibitem[2002]{bohringer02} B{\" o}hringer,
H., Matsushita, K., Churazov, E., Ikebe, Y., \& Chen, Y.\ 2002, \aap,
382, 804

\bibitem[1996]{boulanger96} Boulanger, F. et al. A\&A, 312, 256

\bibitem[2001]{bullock} Bullock, J.S., Kollat, T.S., Sigad, Y. \etal
2001, MNRAS, 321, 559


\bibitem[2003]{buote03} Buote, D.A., \& Lewis, A.D., 2003, ApJ in 
press, astro-ph/0312109s

\bibitem[Chen, Ikebe, \& B{\" o}hringer 2003]{chen03} Chen,
Y., Ikebe, Y., \& B{\" o}hringer, H.\ 2003, \aap, 407, 41

\bibitem[2001]{david01}
              David, L.P., Nulsen, P.E.J., McMamara, B.R., \etal\  2001,
ApJ, 557, 546


\bibitem[De Grandi \& Molendi 1999]{degrandi99} De Grandi, S.~\&
Molendi, S.\ 1999, \aap, 351, L45

\bibitem[1990]{dickey90} Dickey, J.M., Lockman, F.J. 1990, ARA\&A, 28,
215

\bibitem[1991]{edge91} Edge, A.~C.~\&
Stewart, G.~C.\ 1991, \mnras, 252, 414

\bibitem[2001]{edge01} Edge, A.~C.\ 2001, \mnras, 328, 762

\bibitem[Eke, Navarro, \& Frenk 1998]{eke98} Eke, V.~R.,
Navarro, J.~F., \& Frenk, C.~S.\ 1998, \apj, 503, 569

\bibitem[Fabian 1994 ]{fabian94} Fabian, A.~C.\ 1994, \araa, 32,
277

\bibitem[2001]{ghizzardi01} Ghizzardi, S. 2001, EPIC-MCT-TN-011
(XMM-SOC-CAL-TN-0022)

\bibitem[2002]{ghizzardi02} Ghizzardi, S. 2002, EPIC-MCT-TN-012

\bibitem[1997]{hartmann97} Hartmann, D.~\&
Burton, W.~B.\ 1997, ``Atlas of Galactic neutral hydrogen'',
Cambridge; New York: Cambridge University Press, ISBN 0521471117

\bibitem[Hicks et al. 2002]{hicks02}
Hicks, A.~K., Wise, M.~W., Houck, J.~C., \& Canizares, C.~R.\ 2002,
\apj,
580, 763

\bibitem[1992]{johnstone92}
Johnstone, R.~M., Fabian, A.~C., Edge, A.~C., \& Thomas, P.~A.\ 1992,
\mnras, 255, 431


\bibitem[2003]{kaastra03}Kaastra, J.~S., et al.\ 2003, \aa in press,
\astroph0309763

\bibitem[2003]{lewis03} Lewis, A.D., Buote, D.A., Stocke, J.T., 2003,
\apj, 586, 135

\bibitem[2002]{lumb02} Lumb,
D.~H., Warwick, R.~S., Page, M., \& De Luca, A.\ 2002, \aap, 389, 93

\bibitem[2002]{matsushita02} Matsushita, K., Belsole, E., Finoguenov,
A., \& B{\" o}hringer, H.\ 2002, \aap, 386, 77

\bibitem[Markevitch 1998]{markevitch98} Markevitch, M.\ 1998, \apj,
504, 27

\bibitem[2001]{molendi01} Molendi, S.~\&
Pizzolato, F.\ 2001, \apj, 560, 194

\bibitem[Montier \& Giard 2003]{montier03} Montier, L., Giard, M. 2003,
\aa~ accepted

\bibitem[1999]{moore99} Moore, B., Quinn, T.,
Governato, F., Stadel, J., \& Lake, G.\ 1999, \mnras, 310, 1147

\bibitem[1997]{navarro97} Navarro,
J.~F., Frenk, C.~S., \& White, S.~D.~M.\ 1997, \apj, 490, 493

\bibitem[2001]{peterson01} Peterson, J.~R., et al.
2001, \aap, 365, L104

\bibitem[2003]{peterson03} Peterson, et al.
2003, \apj, 590, 207

\bibitem[2002]{pratt02} Pratt, G.~W.~\& Arnaud, M.\ 2002, \aap, 394, 375

\bibitem[2003]{pratt03} Pratt, G.~W.~\& Arnaud, M.\ 2003, \aap, 408, 1

\bibitem[1998]{schlegel98}
Schlegel, D.~J., Finkbeiner, D.~P., \& Davis, M.\ 1998, \apj, 500,
525

\bibitem[Shibai 2002]{shibai02} Shibai, H.\ 2002, Advances in
Space Research, 30, 2089

\bibitem[1997]{snowden97} Snowden S.L., Egger R., Freyberg M. J. \etal,
1997, ApJ, 485, 125

\bibitem[2002]{snowden02} Snowden, S.  2002, in Proceedings of the
"New Vision of
the X-ray Universe in the \xmm  and Chandra Era" conference, to appear

\bibitem[Struble \& Rood 1999]{struble99} Struble, M.~F.~\& Rood, 
H.~J.\ 1999, \apjs, 125, 35

\bibitem[2003]{sun03} Sun, M., Jones, C., Murray,
S.~S., , S.~W., Fabian, A.~C., \& Edge, A.~C.\ 2003, \apj, 587,
619

\bibitem[1994]{white94} White, D.~A., Fabian,
A.~C., Allen , S.~W., \etal\ 1994, \mnras, 269, 589

\bibitem[2000]{white00} White, D.~A.\ 2000, \mnras, 312,
663
\end{thebibliography}
\end{document}